\documentclass[pdftex,twocolumn,epjc3]{svjour3}          

\RequirePackage[T1]{fontenc}

\smartqed  

\RequirePackage{graphicx}
\RequirePackage{mathptmx}      
\RequirePackage{flushend}
\RequirePackage[numbers,sort&compress]{natbib}
\RequirePackage[colorlinks,citecolor=blue,urlcolor=blue,linkcolor=blue]{hyperref}

\journalname{Eur. Phys. J. C}

\usepackage{array}
\usepackage[format=plain,labelformat=default,labelsep=period]{caption}
\usepackage[margin=10pt]{subcaption}
\usepackage{amsmath}
\usepackage{xstring}
\def\ReplaceStr#1{%
  \IfSubStr{#1}{/}{%
    \StrSubstitute{#1}{/}{\slash }}{#1}}

\usepackage{soul}
\usepackage[dvipsnames]{xcolor}
\usepackage{etoolbox}

\newtoggle{HighlightDiff}

\togglefalse{HighlightDiff}

\usepackage{hyphenat}
\begin{document}
\title{Simulating the Time Projection Chamber responses at the MPD detector using Generative Adversarial Networks}

\author{A. Maevskiy\thanksref{e1,addr1}
        \and
        F. Ratnikov\thanksref{e2,addr1,addr2}
        \and
        A. Zinchenko\thanksref{e3,addr3}
        \and
        V. Riabov\thanksref{e4,addr4}
}

\thankstext{e1}{e-mail: artem.maevskiy@cern.ch}
\thankstext{e2}{e-mail: fedor.ratnikov@cern.ch}
\thankstext{e3}{e-mail: alexander.zinchenko@jinr.ru}
\thankstext{e4}{e-mail: riabovvg@gmail.com}

\institute{HSE University, 20 Myasnitskaya Ulitsa, Moscow, Russia\label{addr1}
          \and
          Yandex School of Data Analysis, 11-2 Timura Frunze Street, Moscow, Russia\label{addr2}
          \and
          Joint Institute for Nuclear Research, 6 Joliot-Curie St, Dubna, Moscow Oblast, Russia\label{addr3}
          \and
          Petersburg Nuclear Physics Institute, 1, mkr. Orlova roshcha, Gatchina, Leningradskaya Oblast, Russia\label{addr4}
}

\date{Received: date / Accepted: date}

\iftoggle{HighlightDiff}{%
  \newcommand{\RemoveText}[1]{{\color{RubineRed}\st{#1}}}
  \newcommand{\AddText}[1]{{\color{ForestGreen}#1}}
}{%
  \newcommand{\RemoveText}[1]{}
  \newcommand{\AddText}[1]{#1}
}
\maketitle

\begin{abstract}
High energy physics experiments rely heavily on the detailed detector simulation models in many tasks. Running these detailed models typically requires a notable amount of the computing time available to the experiments. In this work, we demonstrate a\RemoveText{n original}\AddText{~new} approach to speed up the simulation of the Time Projection Chamber tracker of the MPD experiment at the NICA accelerator complex. Our method is based on a Generative Adversarial Network~--- a deep learning technique allowing for implicit estimation of the population distribution for a given set of objects. This approach lets us learn and then sample from the distribution of raw detector responses, conditioned on the parameters of the charged particle tracks. To evaluate the quality of the proposed model, we integrate a prototype into the MPD software stack and demonstrate that it produces high-quality events similar to the detailed simulator, with a speed-up of at least an order of magnitude. The prototype is trained on the responses from the inner part of the detector and, once expanded to the full detector, should be ready for use in physics tasks.
\end{abstract}

\section{Introduction}
\label{sec:intro}

Computer simulations of high-energy physics experiments play a crucial role in a variety of relevant tasks, including detector geometry optimization~\cite{Baranov:2017chy,Boldyrev:2020ydy}, selecting best analysis strategies~\cite{Drnoyan:2020acj,Kolesnikov:2020rai}, and testing the Standard Model (SM) predictions and searching for new phenomena beyond the SM~\cite{Aaboud:2019lgy,Aad:2019vvf}. For a typical experimental data analysis, the number of simulated events usually translates directly to the uncertainty of the final physics result. The amount of computational resources spent on the simulations usually takes a notable fraction of the total computing capabilities of an experiment and is comparable with that spent on the real data processing \cite{Alves:2017she,Chapman:2020nnj}. Therefore, faster approaches to event generation and simulation are in great demand for the existing and future high energy physics experiments.

The MPD detector is one of the two experiments at the NICA accelerator complex~--- a new heavy ion accelerator facility being constructed at the Joint Institute for Nuclear Research and located in Dubna, Russia~\cite{Trubnikov:2010zz,Abraamyan:2011zz}. The complex is designed to study the properties of dense baryonic matter. For the tracking, MPD utilizes a time projection chamber (TPC) in the central barrel~\cite{Averyanov:2020lgp}. TPC simulation is very CPU-intensive~\cite{Zinchenko:2018}, and hence a fast simulation approach for TPC is highly desirable.

A typical approach to constructing models for fast simulation of particle physics detectors is to use a simplified detector geometry and a simplified model of the interaction of particles with matter~\cite{Lukas:2012kua}. This approach is justified for subsystems with a flat sensitive volume, such as silicon trackers, that measure the two-dimensional coordinate of a passing particle. For systems with a large volume, such as calorimeters or TPC-based trackers, this approach makes it difficult to achieve a reasonable compromise between accuracy and simulation speed.

Another fast simulation approach is an analytical parameterization of the detector responses, as can be seen in shower shape parameterizations for calorimeters~\cite{Yamanaka:2011zz}. This approach can significantly speed up the calorimeter simulation, but it makes it difficult to achieve high quality simulated data. A common solution for calorimeters is also to use the so-called "frozen showers"~\cite{Lukas:2012kua} when detailed simulated system responses are stored as a response library for subsequent reuse.
 
With recent developments and particular success of deep learning in high energy physics~\cite{deOliveira:2015xxd,Aurisano:2016jvx,Andrews:2018nwy,Andrews:2019faz,DiBello:2020bas}, fast simulation models based on a Generative Adversarial Network (GAN) \cite{Goodfellow:2014upx} and Variational Auto-Encoders (VAE) \cite{rezende2014stochastic,kingma2014autoencoding} have emerged (see e.g.~\cite{deOliveira:2017pjk,Paganini:2017hrr,Erdmann:2018kuh,Erdmann:2018jxd,Chekalina:2018hxi,ATLAS:FastShowerSim2018,Musella:2018rdi,Derkach:2019qfk,Maevskiy:2019vwj,Deja:2019vcv,DiSipio:2019imz,Chapman:2020nnj,Diefenbacher:2020rna,Hariri:2021clz}). These techniques allow to learn the data distribution, with sampling being as fast as a single forward pass through the neural network. In this work, we focus on using GANs for fast simulation. GAN training is based on a competition between two independent neural networks — the generator and discriminator. The generator aims to convert samples from a fixed known distribution into the objects from the target distribution,~--- the one that the training data follows. The discriminator network takes the examples from the target and generator output distributions and predicts the probability for each of these examples to belong to the target distribution. The objective of the discriminator is to \AddText{learn a metric that~}maximize\AddText{s} the separation between the training data and the generated objects, while that of the generator is to minimize this separation. In~\cite{Goodfellow:2014upx} it is shown that the equilibrium state of such a system is a situation where the objects generated by the generator are indistinguishable from those of the training sample, and hence the generator has learned the training data distribution.

In this work, we propose a new method for fast simulation of TPC trackers applied to the tracker of the MPD detector at the NICA accelerator complex, using a GAN.

The structure of the paper is the following. In section~\ref{sec:relatedwork}, previous research in applying deep generative models to the fast simulation of particle physics detectors is discussed. In section~\ref{sec:tpcmpd}, we describe the TPC detector at the MPD experiment. Section~\ref{sec:modeldescription} explains our approach to simulation of the TPC with GANs. In section~\ref{sec:resultsandvalidation}, we demonstrate our results and evaluate their quality. Section~\ref{sec:discussion} is dedicated to the discussion of the perspectives and limitations of our approach. Finally, section~\ref{sec:conclusion} contains a short summary of the work.

\section{Related work}
\label{sec:relatedwork}
As was discussed in section~\ref{sec:intro}, GANs have previously been applied as a tool for fast simulation of high energy physics experiments~\cite{deOliveira:2017pjk,Paganini:2017hrr,Erdmann:2018kuh,Erdmann:2018jxd,Chekalina:2018hxi,ATLAS:FastShowerSim2018,Musella:2018rdi,Derkach:2019qfk,Maevskiy:2019vwj,Deja:2019vcv,DiSipio:2019imz,Chapman:2020nnj,Diefenbacher:2020rna,Hariri:2021clz}. This idea is first proposed in~\cite{deOliveira:2017pjk}, and then further developed in~\cite{Paganini:2017hrr}, where it is shown that GANs can significantly speed up electromagnetic calorimeter simulation by generating raw calorimeter readouts. In~\cite{Chekalina:2018hxi}, this approach is further developed for the case of the LHCb calorimeter, and in~\cite{ATLAS:FastShowerSim2018}~--- for the ATLAS calorimeter. Calorimeters from high energy physics experiments are particularly appealing to simulate using GANs. Their responses form 2- or 3-dimensional structures of fixed size, which are similar to regular images, where GANs have demonstrated outstanding performance over the past years~\cite{gui2020review}. Use of GANs for simulation of the reconstructed characteristics is proposed for the case of Cherenkov detectors in~\cite{Derkach:2019qfk}, and further developed for LHCb RICH detectors in~\cite{Maevskiy:2019vwj}. Simulation of reconstructed objects has also been proposed in \cite{Martinez:2019jlu}. In \cite{Musella:2018rdi}, GANs are used for simulating hadronic jets, as simulated and reconstructed in the CMS detector. This research is not limited to the collider experiments alone, however. E.g. in~\cite{Erdmann:2018kuh} GANs are used for simulating a ground-based array of particle detectors for cosmic rays.

Using GANs for the fast simulation of the TPC type detector has been studied for the ALICE experiment~\cite{Deja:2019vcv}. The main idea in that work is to use a GAN to model the measured cluster coordinates, conditioned on the charged particle track parameters. With such an approach, the size of the target representation, i.e. the number of measured track interaction points, varies from track to track. This is problematic to model with simple feed-forward or convolutional neural network architectures and typically requires more computationally expensive and harder to train techniques. The authors of~\cite{Deja:2019vcv} overcome this problem by fixing the size of the target representation to the maximum possible number of points per track and zero-padding the trajectories that have fewer points. The quality of the generated data is measured by the mean squared distance between the generated points and the true track helix. It is shown, that the proposed model cannot yet match the quality of the training data, although it accelerates the detailed simulation by a factor of 25~\cite{Deja:2019vcv}.

The idea of utilizing GANs for simulating TPC response at the MPD detector has previously been proposed in~\cite{Zinchenko:2018}. In our work, we develop upon that concept to model the raw TPC readout electronics response, rather than the measured track coordinates as done in~\cite{Deja:2019vcv}. This allows us to utilize the translational symmetries of the detector to reduce the dimensionality of the learned representation, and also to avoid the problem of the variable number of points per track. Another difference from~\cite{Deja:2019vcv} is that we condition the generation process on the parameters of the small track segments that contribute to the simulated response, rather than the parameters of the track at the production point. This means that our method can be used for events with any topology, including e.g. the effects of decay-in-flight, when a particle decays within the detector volume. This is opposed to the approach from \cite{Deja:2019vcv} where only long enough tracks are considered and taking such effects into account would require training the model specifically for such events.

We would like to finalise this section with a short note on the applicability of GAN usage for fast detector simulation. As it is raised in \cite{Matchev:2020tbw}, for a physics analysis, the systematic uncertainties associated with a GAN-generated sample cannot be smaller compared to those from the sample that GAN is trained on. \AddText{At the same time, applying GANs also means making use of the prior knowledge about the structure of the data \cite{Butter:2020qhk}.} In other words, the generated data does not contain more information about the true distribution than there is in the training sample \AddText{combined with the prior knowledge introduced by the network architecture}. One should note, that the same argument applies to other fast simulation techniques like analytical parameterizations or memorized event libraries. Our position is that deep generative modeling should rather be considered as a method for memorization and interpolation of the training data, and it should be applicable in those cases where analytical parameterizations and memorized event libraries are.

\section{TPC tracker of the MPD experiment}
\label{sec:tpcmpd}
TPC is the main tracking detector of the central barrel of the MPD experiment~\cite{TPCTDR:2019v07}, covering the pseudorapidity region of $\left|\eta\right| < 1.2$ and $2\pi$ in azimuthal angle. It is designed to provide the high efficiency of the charged particle track reconstruction with a momentum resolution of better than 3\% in the transverse momentum range $0.1 < p_\text{T} < 1$\,GeV/c. The double-track resolution should not exceed 1\,cm for operation in a high-multiplicity environment realized in central collisions of heavy ions. Moreover, the TPC is used as one of the primary particle identification detectors. The particle identification is based on the measurement of the ionization losses with a resolution better than 8\%.

The gas volume of the detector occupies the central barrel radial region between  34 and 134\,cm and has a length of 3.4\,m in the beam direction (z-axis). The detector is aligned along the beam axis and is centered with respect to the nominal collision point. The gas volume of the detector is enclosed in an electric field cage, which together with the central membrane effectively divides it into two identical halves with the electric field lines parallel to the beam axis. The primary ionization electrons drift towards the edges, and the signals are amplified and registered with the proportional chambers with the cathode readout. The chambers are arranged in 12 sectors on each side of the detector with each sector covering 30 degrees in azimuth. The cathode planes of the chambers are pad-segmented with 53 pad rows perpendicular to the radial direction and the number of pads in each row increasing with the radius. The pads are 5mm wide, while their height is 12\,mm (short pads) and 18\,mm (long pads) for the 27 inner and 26 outer rows, respectively. Overall, the detector has 95232 sensitive pads, responses from which are collected in 310 time buckets per bunch crossing. The XY-coordinates of the track segments are obtained from the location of the fired pads. The z-coordinate is calculated based on the measured drift time and the known drift velocity of electrons in the gas.

\section{Model description}
\label{sec:modeldescription}

As was mentioned in section~\ref{sec:relatedwork}, we build our model to generate raw TPC responses, i.e. the responses from the sensitive pads. Given the total number of pads in TPC and the number of time buckets per event their response is recorded in, this means that we need to generate almost 30~million numbers per bunch crossing. To reduce the number of dimensions of the target space, we split the pads into smaller conditionally independent subsets, with conditions imposed by the track parameters, as described below.

In order to apply this reduction, the major assumption we make is that the response at a particular pad row depends only on small segments of the particle trajectories, formed by tracks crossing the corresponding \emph{pad plane}. The \emph{pad plane} is the volume swept by the electric field lines  directed at the given pad row. With this assumption, we ignore the effects of electron drift diffusion in the direction orthogonal to the pad plane or the spread of the induced signal over several pad rows. To fix the dimensionality of the input space, we model the contributions from different track segments separately, combining them in an additive manner and ignoring any nonlinear effects. Later we show that the assumptions described here do not significantly affect the quality of the generated events.

Finally, we utilize the fact that responses from a particular track segment are localized in space and time, and therefore we only model a small number of pads and time buckets for a given track segment. Since the pads are identical, we only train our model on the responses from a small subset of pads in a single pad row and translate the predictions onto all other pads. In fact, there are two shapes of pads (short and long), so a more precise simulation of their responses requires either two separate models or an input condition variable specifying the type of the pad to model the response for. In this work, we train our model only on the responses from the short pads.

\subsection{Data}
\label{sec:trainingdata}
The training dataset is obtained by running the detailed MPD simulator~\cite{Kolesnikov:2019psa}, which is based on Geant3 transport of particles through the detector materials and realistic simulation of the detector responses based on the first principles. Each of the simulated events has only one positively charged pion with the transverse momentum of $p_\text{T}=478.3$\,MeV/c. The pions are generated uniformly along the drift path with the azimuthal angle $\varphi$ in a range $[-20; 20]$ degrees and the polar angle $\theta$ in a range $[30, 150]$ degrees. The generated responses are picked up from the 20th pad row for further analysis.

In the coordinate system local to a given sector of the detector, the track segment crossing a particular pad plane is defined by four parameters:
\begin{itemize}
    \item \emph{crossing angle}~--- the angle between the transverse projection of the particle momentum and the normal to the pad plane;
    \item \emph{dip angle}~--- the angle between the full momentum and its transverse projection;
    \item \emph{drift length}~--- distance from the center of the segment to the triggered pad row, measured in the number of time buckets from the bunch crossing to the pad response generation;
    \item \emph{pad coordinate}~--- coordinate along the pad row direction of the projection of the track segment center onto the triggered pad row, measured in pad widths.
\end{itemize}
For the sector selected for training, the crossing angle equals the azimuthal angle $\varphi$, and the dip angle equals to $(\theta - 90)$\,deg. Overall, 20000 segment responses were generated, which were split into training and validation subsets as 75 : 25, respectively.

\subsection{Data preprocessing}

As was mentioned earlier, pad response contributions from a single track segment affect only a few pads and time buckets. For each segment, in order to make use of this response localization, we shift the responses by the integer parts of the drift length and pad coordinate along the time and pad row directions, respectively. After having done this, the responses in the whole training set fit onto a matrix of 8 pads by 16 time buckets, which constitutes our target space. The responses span over several orders of magnitude, so we scale them with $\log_{10}(x + 1)$ for smoother learning.

Since we utilize the invariance of the responses under translations along the transverse plane (and along the pad row direction in particular), it is sufficient to only feed the fractional part of the pad coordinate into our model, rather than the full pad coordinate. As for the drift length, the response characteristics do depend on its absolute value (e.g. the spread of the response in time buckets increases with the drift length due to diffusion effects), so both the fractional part and the full number are fed into the model as two separate features. Providing the fractional part as a separate input is motivated by the fact that the invariance under longitudinal translations does hold to some degree, and, therefore, this additional feature contains meaningful information. Before being fed into the model, the angles and the drift length are linearly scaled down to a $[-1, 1]$ region.

\subsection{Network architecture and the objective function}
\label{sec:networkarchitecture}

\begin{figure}
\begin{minipage}{\columnwidth}
\centering
\includegraphics[width=0.98\textwidth,trim=110 750 90 140,clip]{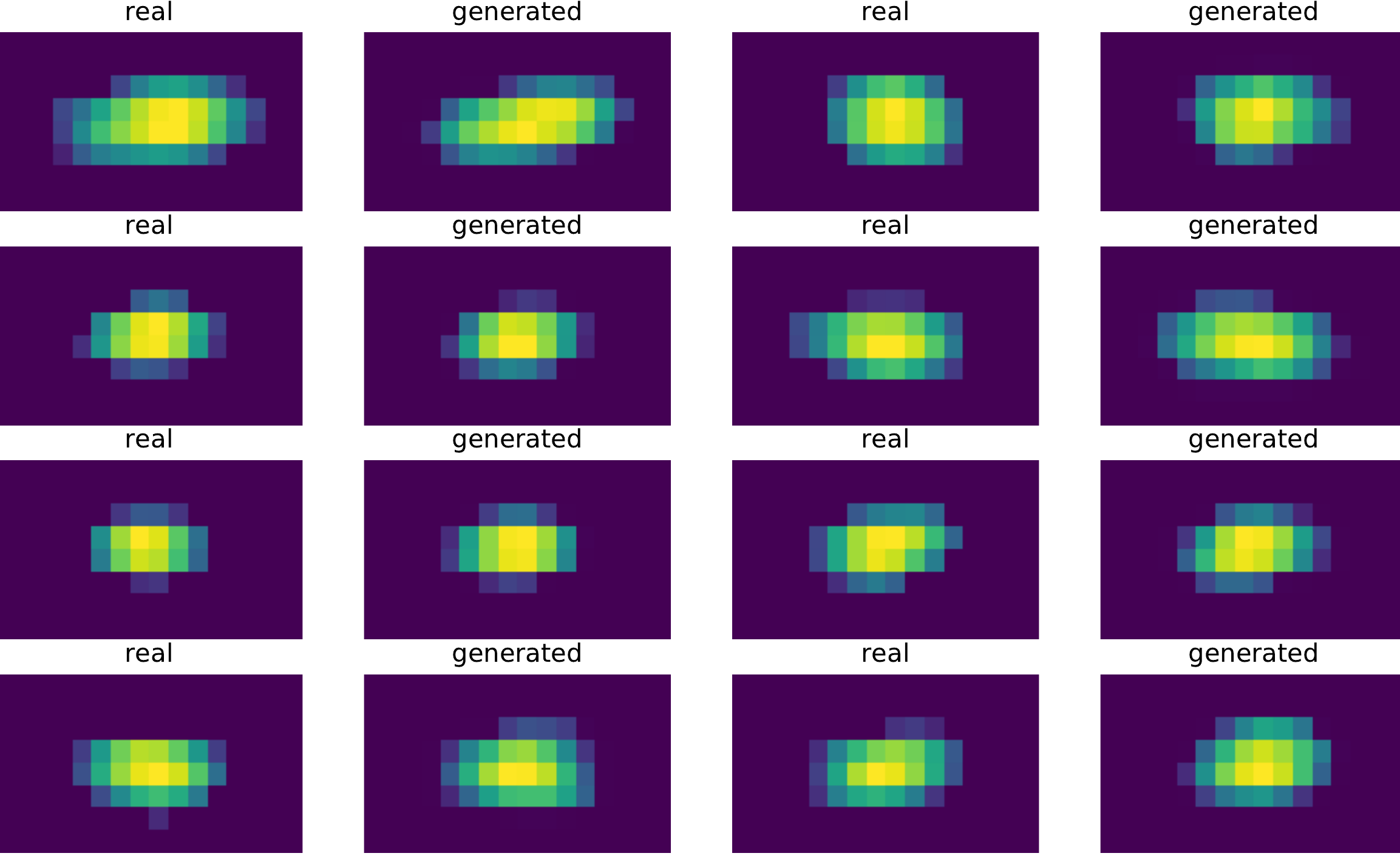}
\end{minipage}
\caption{\label{fig:examplesignals} Examples of the generated pad responses. Vertical and horizontal axes correspond to the pad and time bins, respectively. Each image from the validation dataset (1st and 3rd columns) is paired up with a generated image (2nd and 4th columns) obtained for the same values of the conditional variables.}
\end{figure}

Since our target space has image-like structure, as shown in Fig.~\ref{fig:examplesignals}, it is natural to apply convolutional neural network architectures. However, we obtain that, for our data, the same generated data quality can be achieved with a fully-connected generator architecture that runs much faster compared to a convolutional network (on a single CPU). Therefore, we decide to only use convolutions in the discriminator network, where high performance is not crucial, while switch to a fully-connected network for the generator. Overall, the structures of both networks are given in~\ref{sec:appendix:nnarchitecture}.

In order to provide the features to the discriminator, we tile each of them onto 8x16 matrices to concatenate with the main pad response image along the channels axis. Additionally, we concatenate the features vector to the dense representation obtained at the end of the convolutional part of the network. Alternative architectures, where only one of these two feature paths is kept, demonstrate inferior quality of the generated samples.

\begin{figure}
\begin{minipage}{\columnwidth}
\centering
\includegraphics[width=0.98\textwidth]{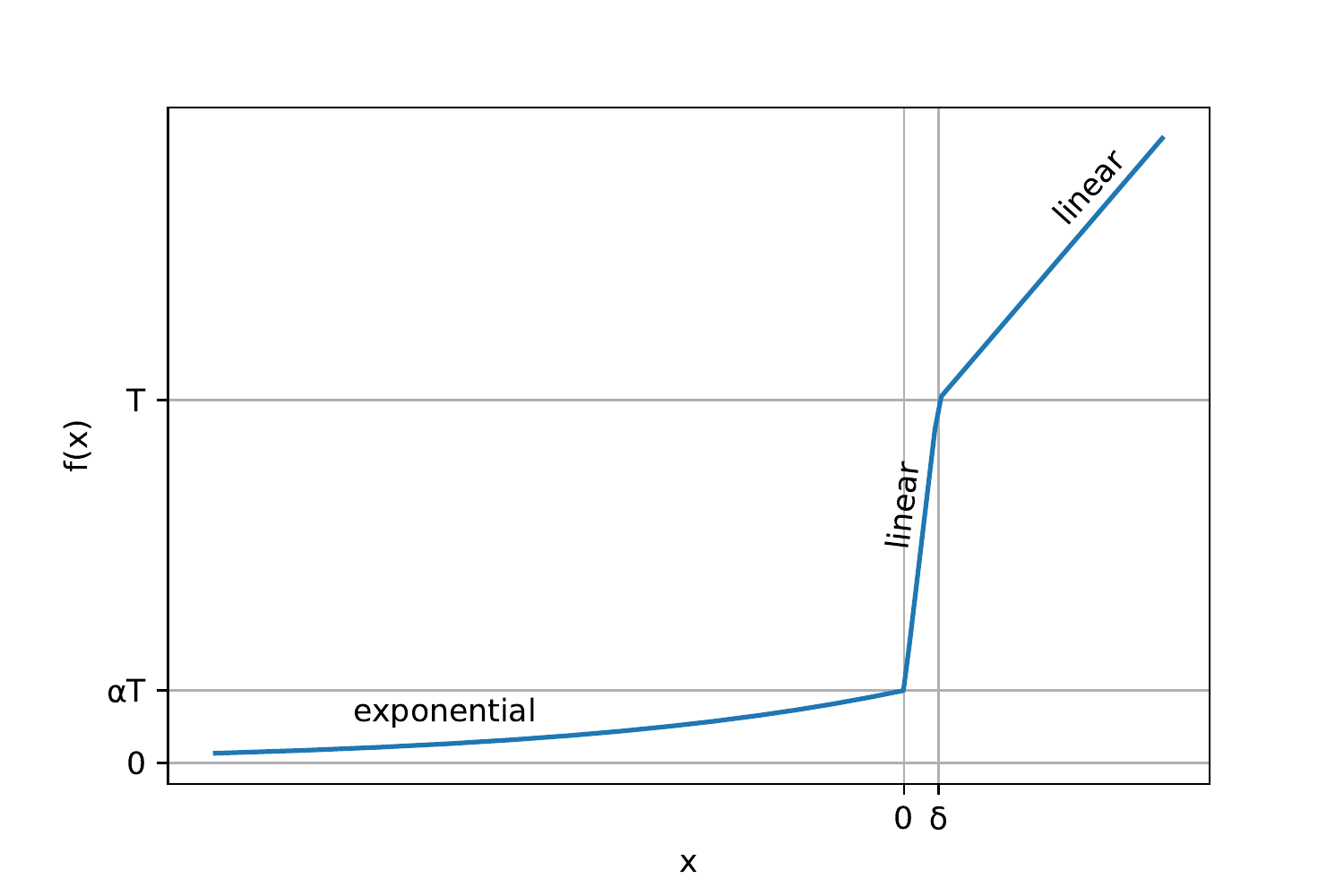}
\end{minipage}
\caption{\label{fig:activation} Activation function $f(x)$ as defined in eq.~(\ref{eq:activation}).}
\end{figure}

In the detailed simulator, the spectrum of the individual pad responses is continuous only down to a certain noise threshold level, below which everything is set to be exactly 0. This means that the target space we are simulating with a GAN contains a discrete mode corresponding to these zero values, which is problematic to learn. In fact, we observe that even the quality of the samples generated above the threshold deteriorates. We attribute this effect to the GAN attempting and failing to describe the steep cutoff with its continuous output. To mitigate this problem, we use a custom activation function at the generator output layer. The function has a very steep slope for the values mapped to the interval between 0 and the threshold, by which we effectively reduce the probability of outputs to end up in that region. The particular form of the function is given by the formula\AddText{~(see Fig.~\ref{fig:activation})}:
\begin{equation}
\label{eq:activation}
f(x) = \begin{cases}
\alpha Te^x     & x\le 0 \\
T\left(\alpha + (1-\alpha)\frac{x}{\delta}\right) & 0\le x\le \delta \\
T - \delta + x & \delta\le x,
\end{cases}
\end{equation}
where $T=\log_{10}2$ is the threshold, $\alpha=0.1$ and $\delta=0.01$. We observe that using this activation function at the output layer of the generator improves the overall quality of the generated samples.

For the GAN objective, we use the Wasserstein distance \cite{arjovsky2017wasserstein} with the gradient penalty term from~\cite{gulrajani2017improved}, as we find it resulting in the best generated data quality. We train both generator and discriminator using RMSprop optimizer with learning rates starting at 0.0001 at the beginning of the training process and decreasing by a factor of 0.999 after each epoch\footnote{By the term \emph{epoch}, we mean a single full pass through the training dataset. When making $k$ discriminator updates per single generator update, this implies that, per epoch, the generator is trained on $\frac{1}{k + 1}$-th fraction of the training data, while the discriminator~--- on the remaining $\frac{k}{k+1}$-th fraction.}. We make 8 discriminator update steps per single generator step. The batch size is 32. \AddText{Model design and training is done using the TensorFlow~2.1 framework~\cite{Abadi:2016kic}.}

\section{Results and validation}
\label{sec:resultsandvalidation}
For visual evaluation of the model, in Fig.~\ref{fig:examplesignals}, we show the example pad response images from the validation dataset paired up with the ones generated with the GAN for the same track segment parameters. As can be seen from the plot, the GAN-generated images are visually similar to those obtained from the detailed simulator.

\begin{figure*}[tbp]
\centering
\begingroup
\tiny
\setlength{\tabcolsep}{0pt}
\begin{tabular}{>{\raggedleft}m{.035\textwidth}>{\centering}m{.19\textwidth}>{\centering}m{.19\textwidth}>{\centering}m{.19\textwidth}>{\centering}m{.19\textwidth}>{\centering\arraybackslash}m{.19\textwidth}}
{\rotatebox[origin=c]{90}{\parbox{.18\textwidth}{\centering Pad barycenter}}} &
\raisebox{-.5\height}{\includegraphics[width=.18\textwidth,trim=0 10 35 30,clip]{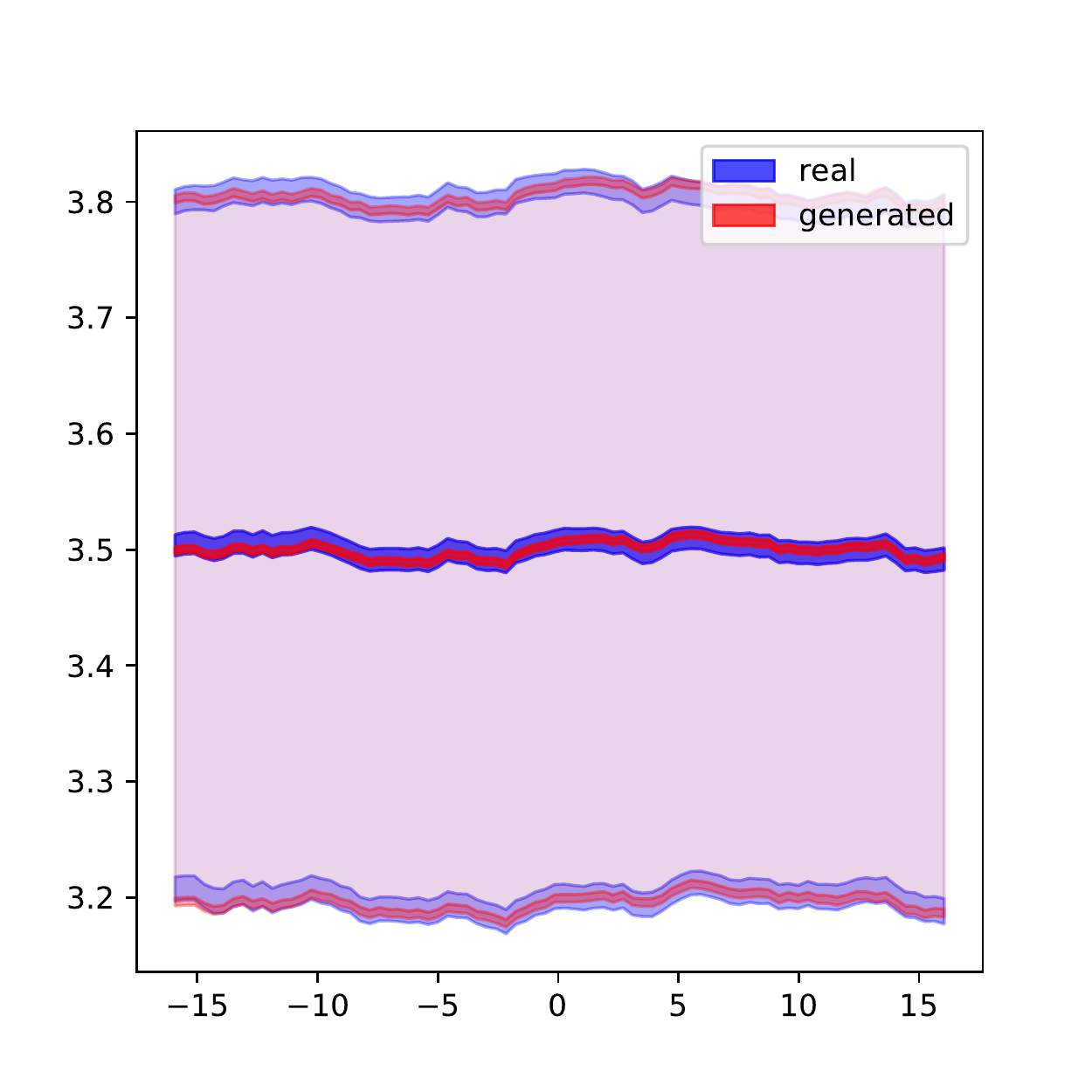}} &
\raisebox{-.5\height}{\includegraphics[width=.18\textwidth,trim=0 10 35 30,clip]{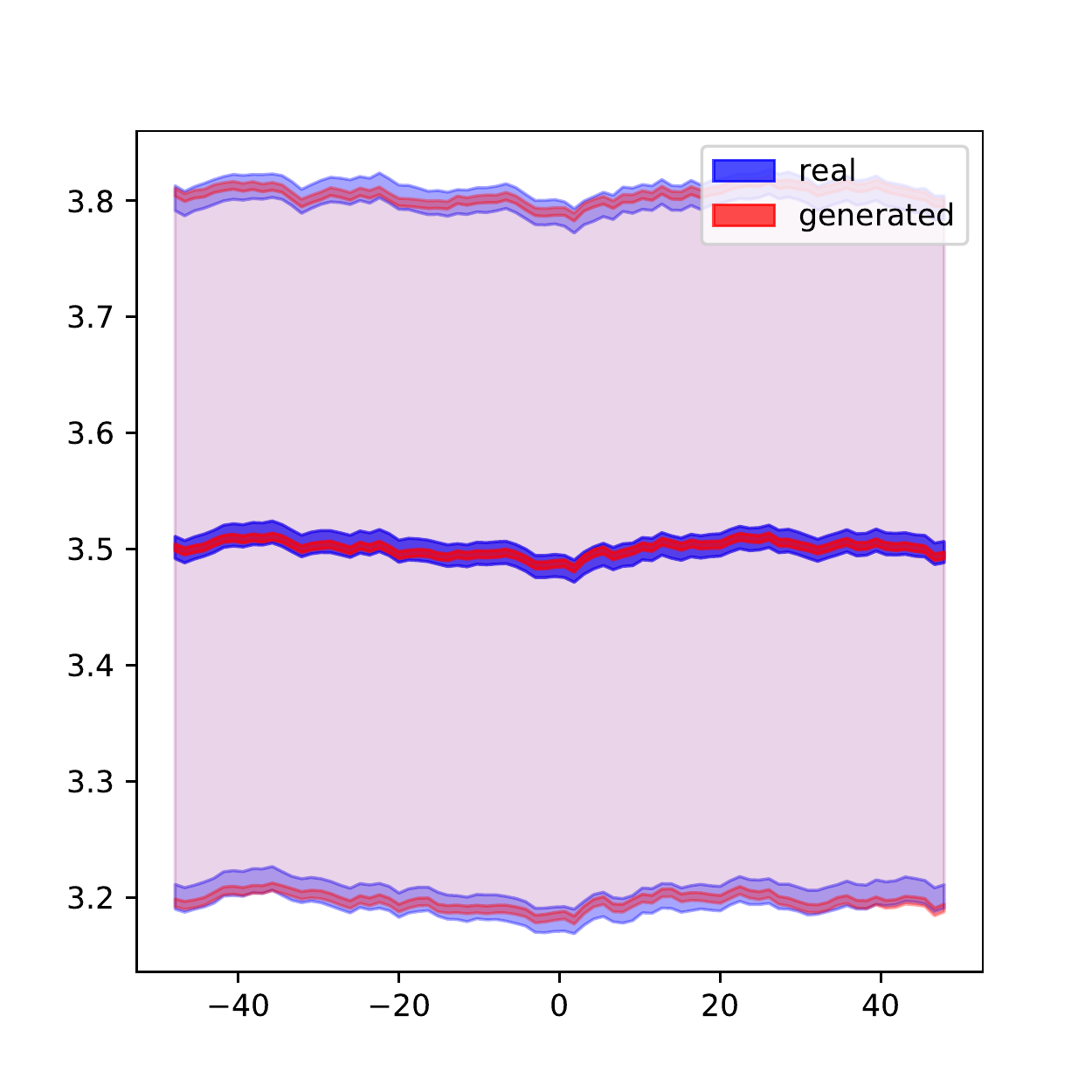}} &
\raisebox{-.5\height}{\includegraphics[width=.18\textwidth,trim=0 10 35 30,clip]{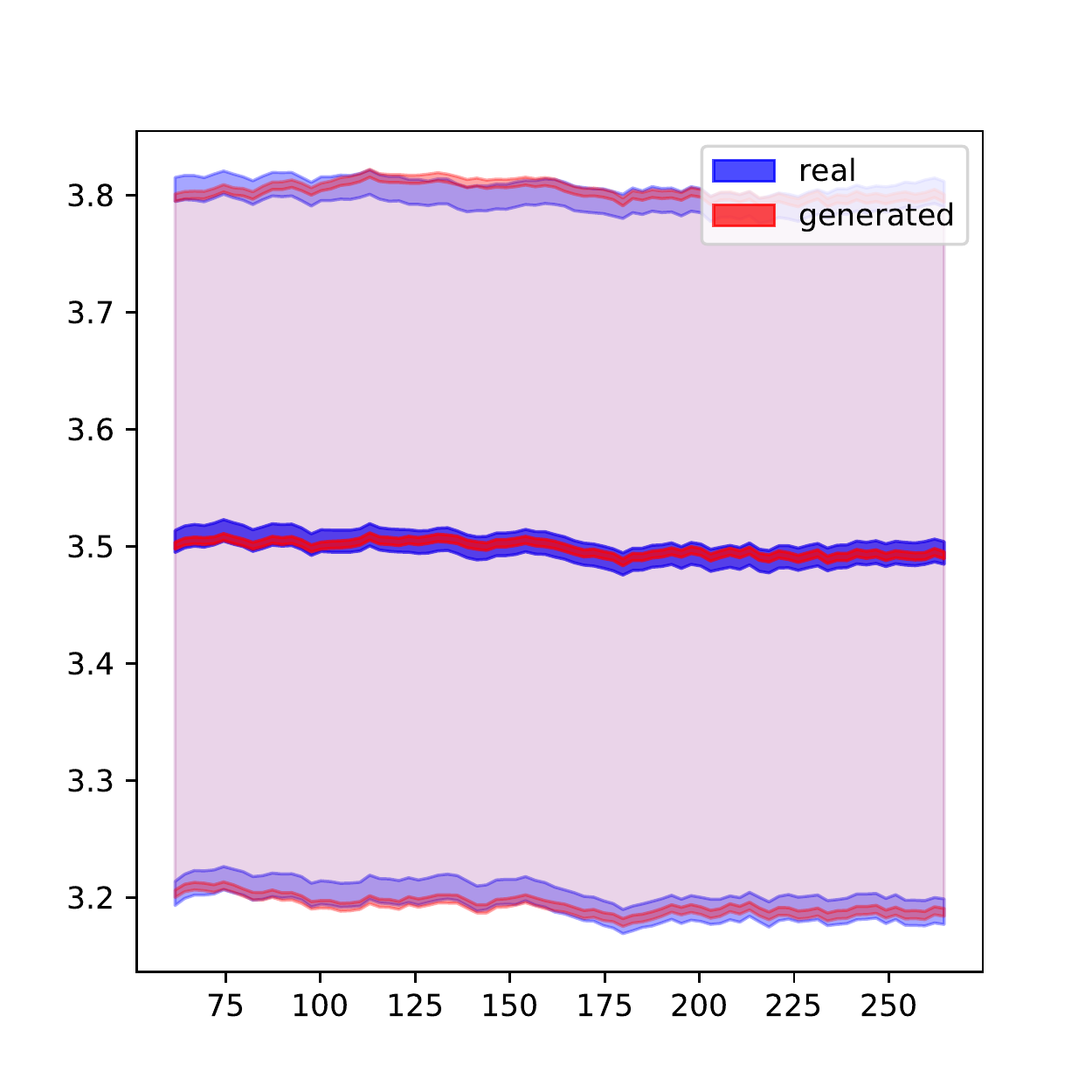}} &
\raisebox{-.5\height}{\includegraphics[width=.18\textwidth,trim=0 10 35 30,clip]{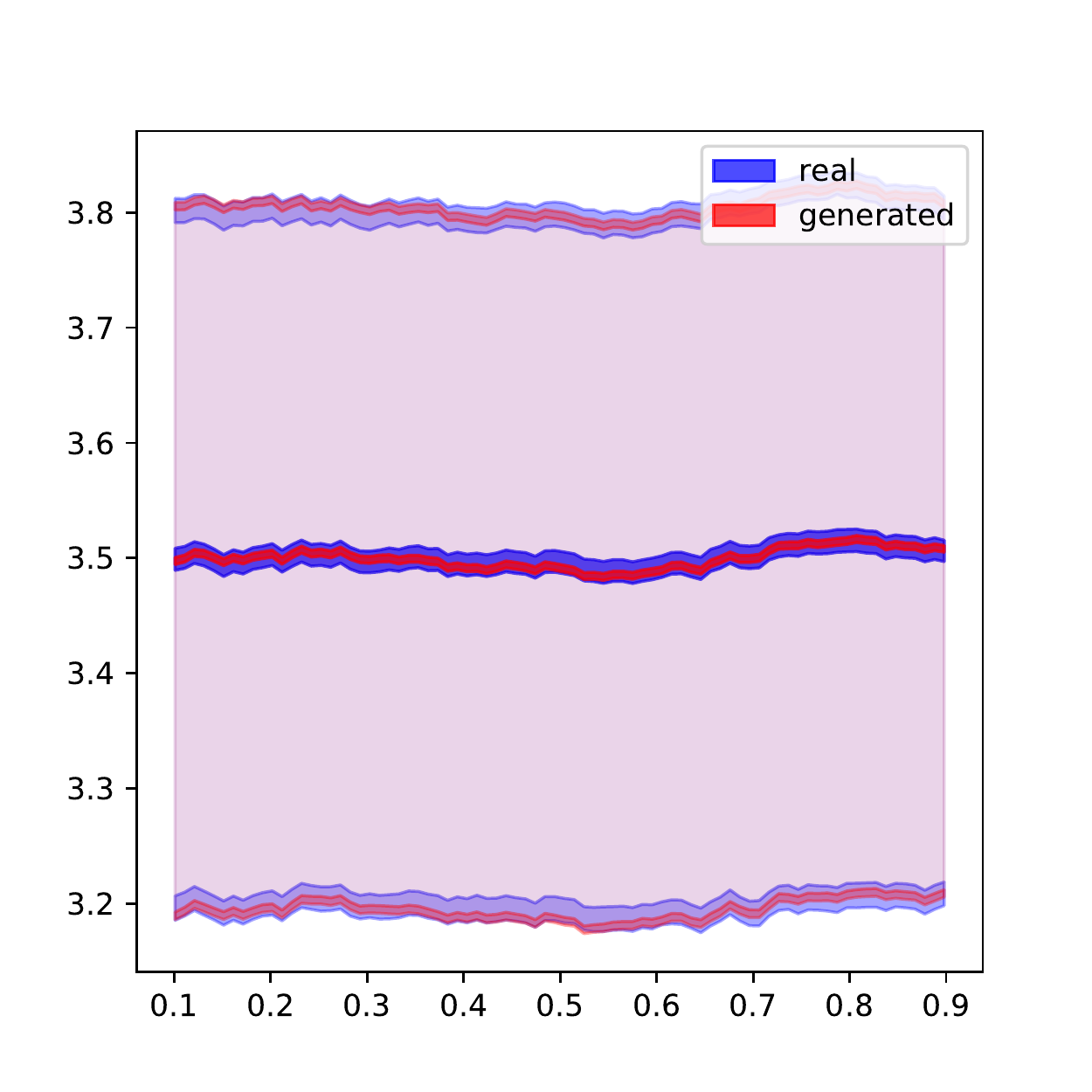}} &
\raisebox{-.5\height}{\includegraphics[width=.18\textwidth,trim=0 10 35 30,clip]{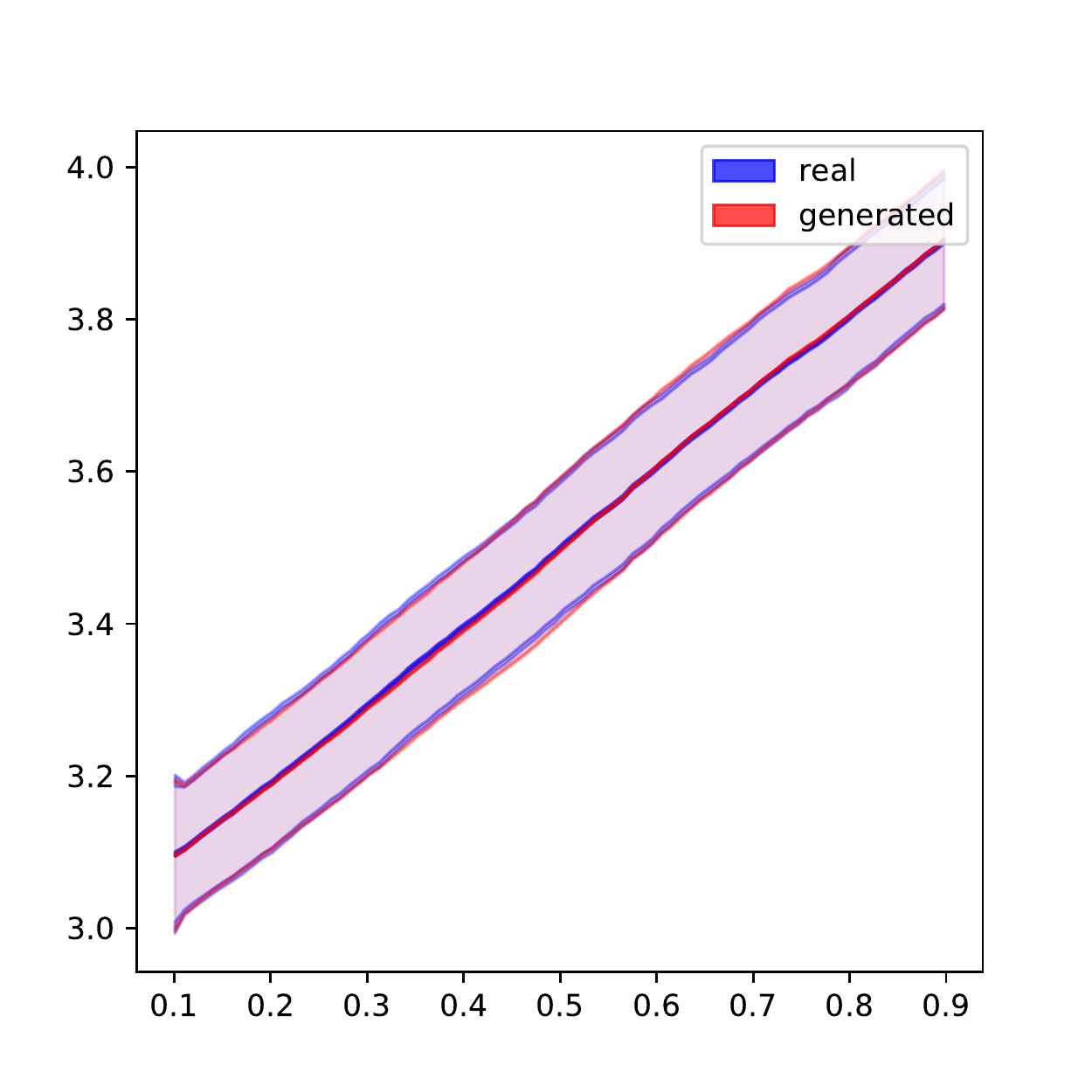}} \\ {\rotatebox[origin=c]{90}{\parbox{.18\textwidth}{\centering Time barycenter}}} &
\raisebox{-.5\height}{\includegraphics[width=.18\textwidth,trim=0 10 35 30,clip]{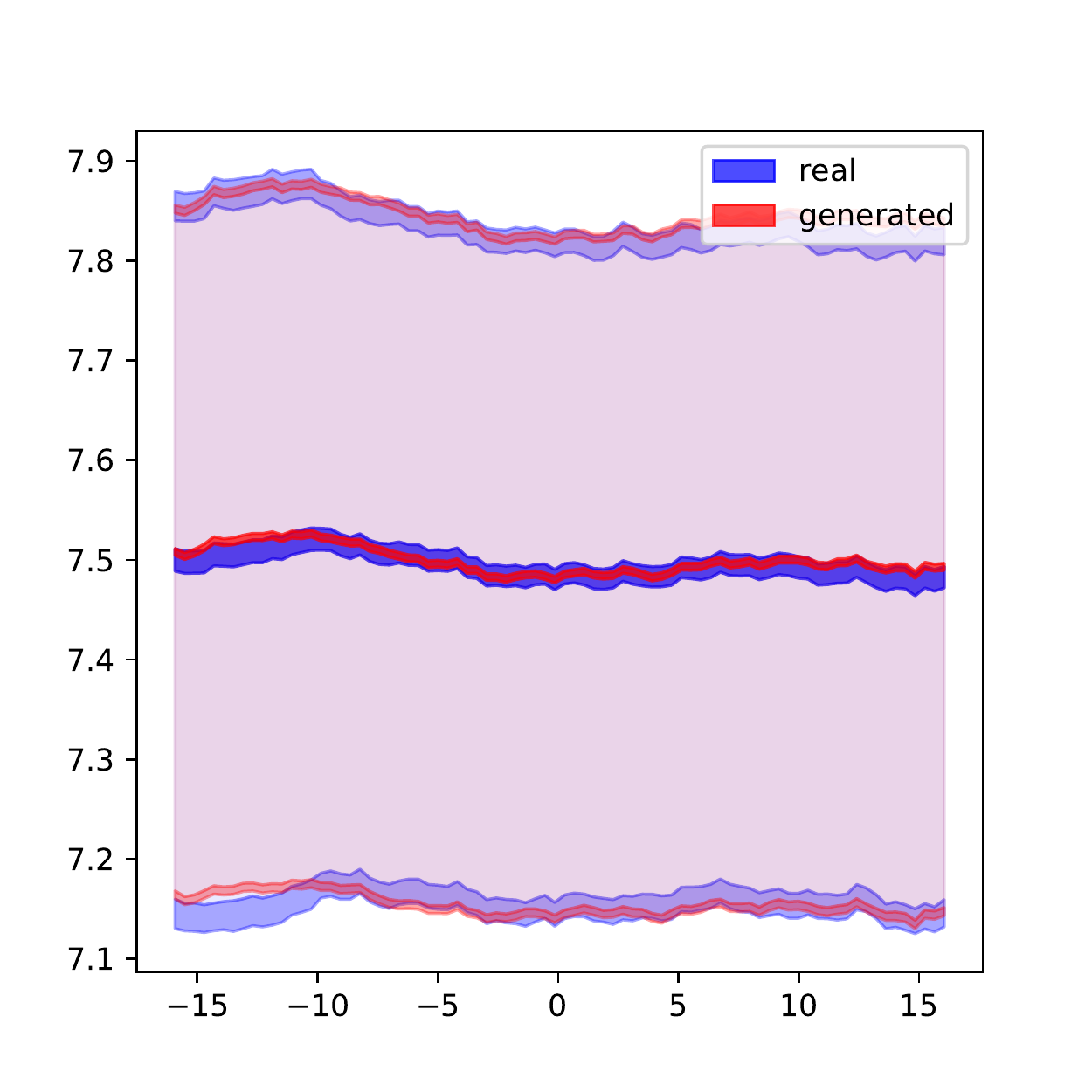}} &
\raisebox{-.5\height}{\includegraphics[width=.18\textwidth,trim=0 10 35 30,clip]{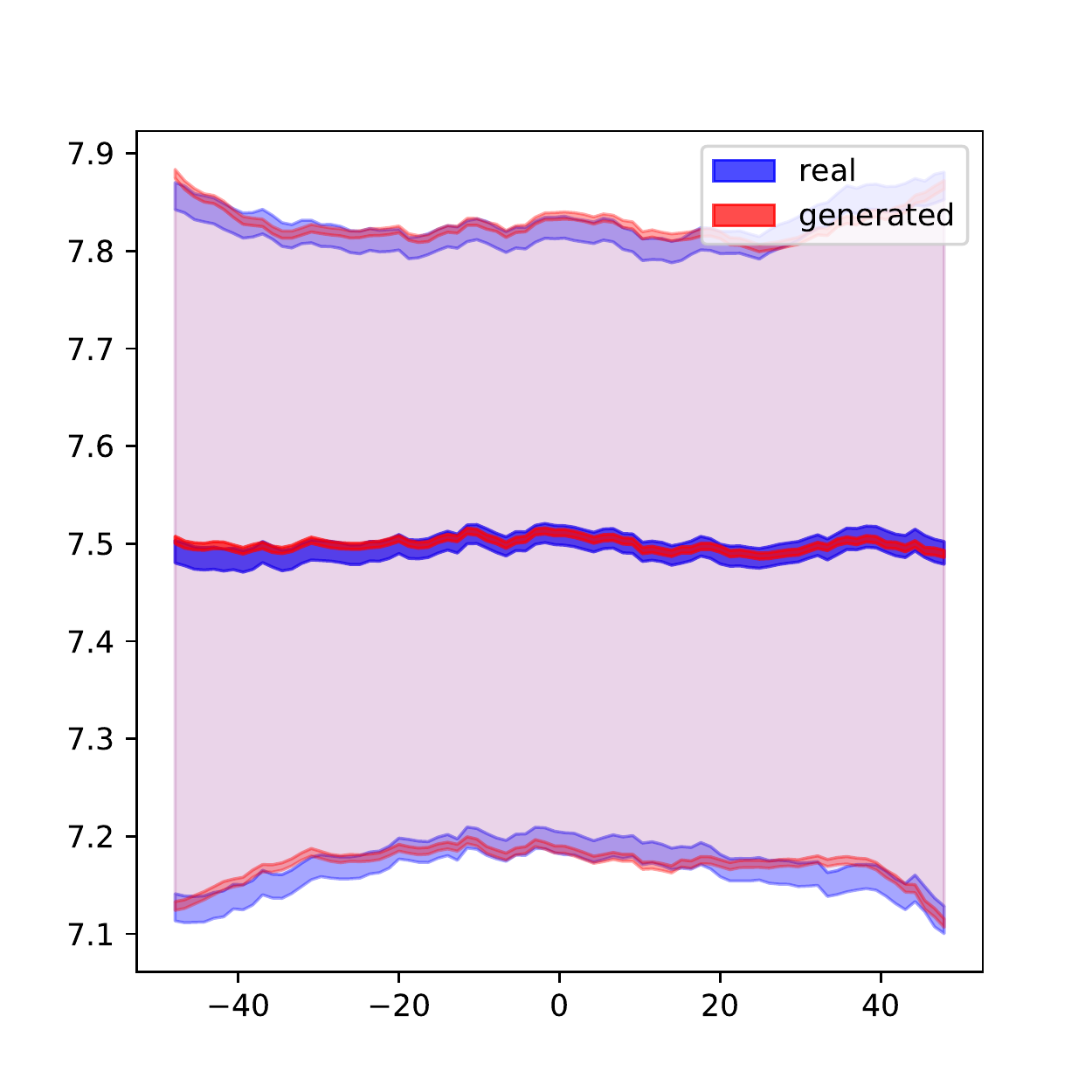}} &
\raisebox{-.5\height}{\includegraphics[width=.18\textwidth,trim=0 10 35 30,clip]{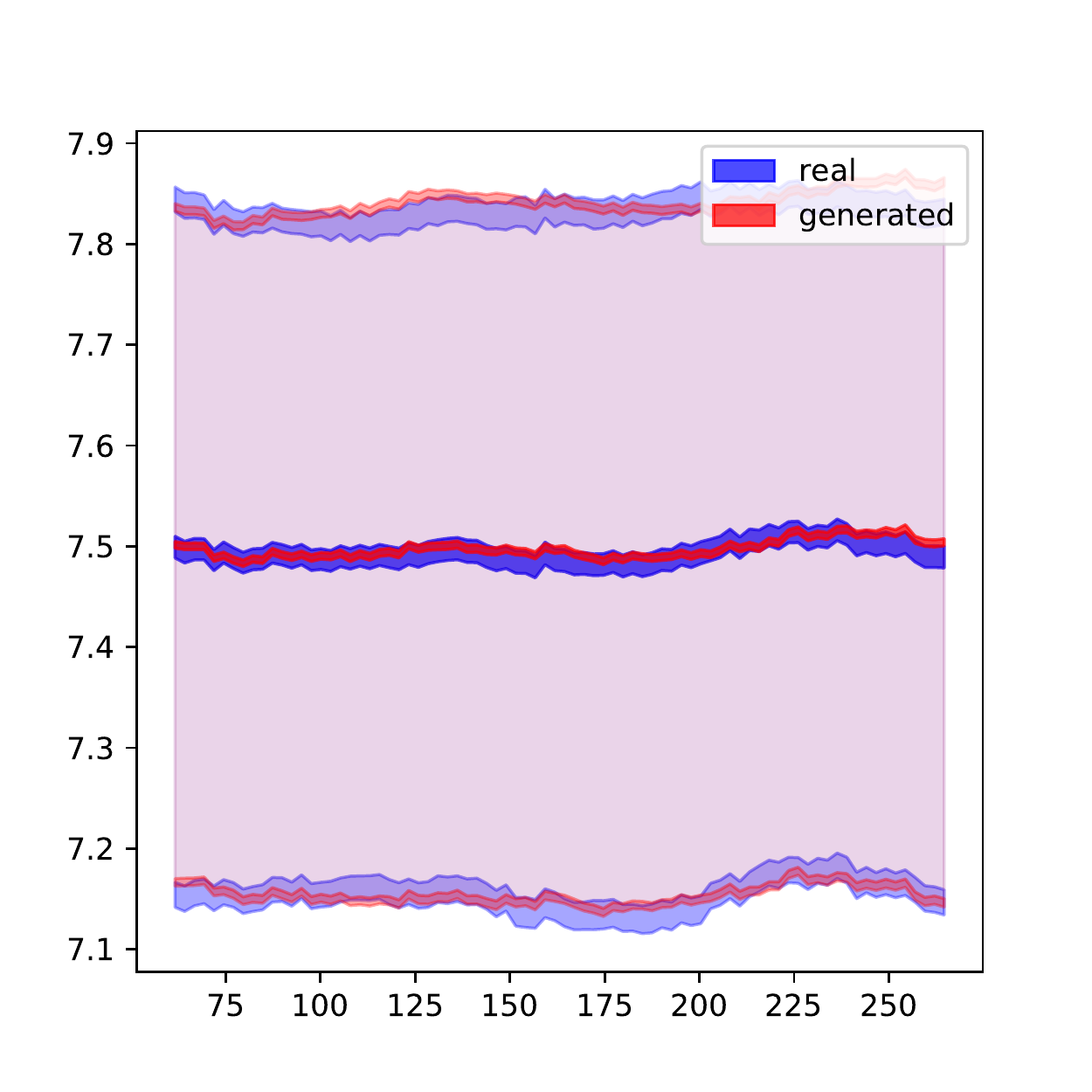}} &
\raisebox{-.5\height}{\includegraphics[width=.18\textwidth,trim=0 10 35 30,clip]{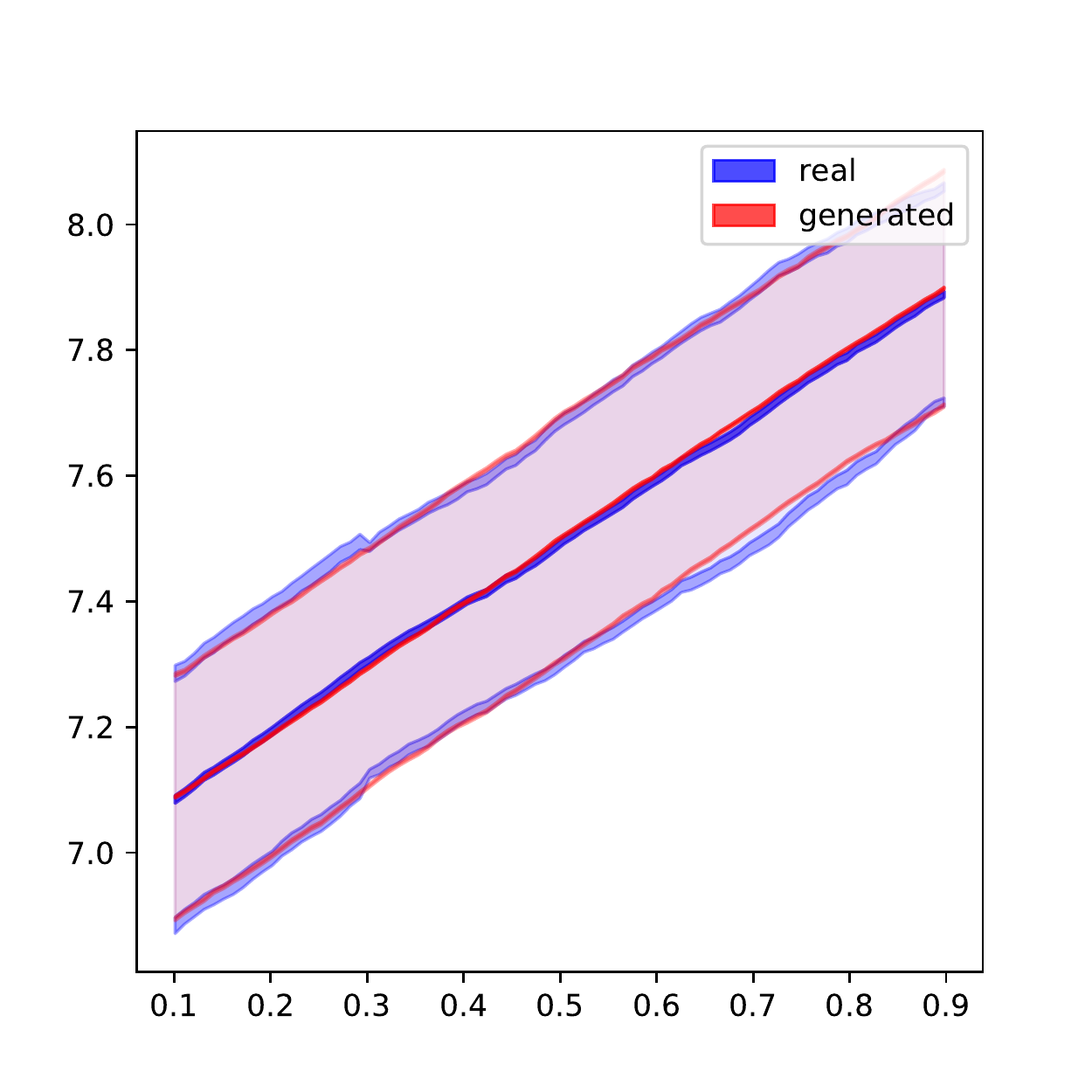}} &
\raisebox{-.5\height}{\includegraphics[width=.18\textwidth,trim=0 10 35 30,clip]{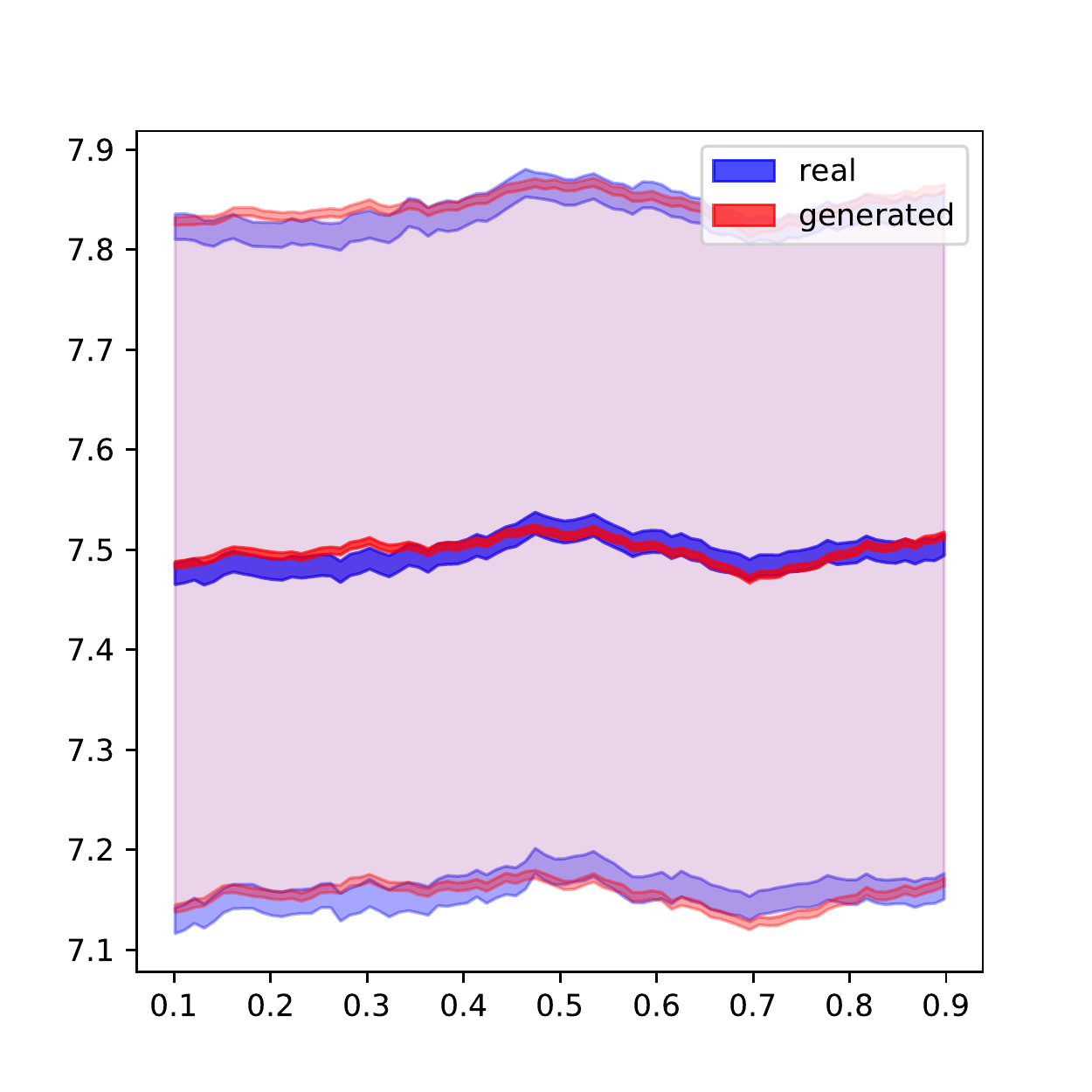}} \\ {\rotatebox[origin=c]{90}{\parbox{.18\textwidth}{\centering Sq. Pad Width}}} &
\raisebox{-.5\height}{\includegraphics[width=.18\textwidth,trim=0 10 35 30,clip]{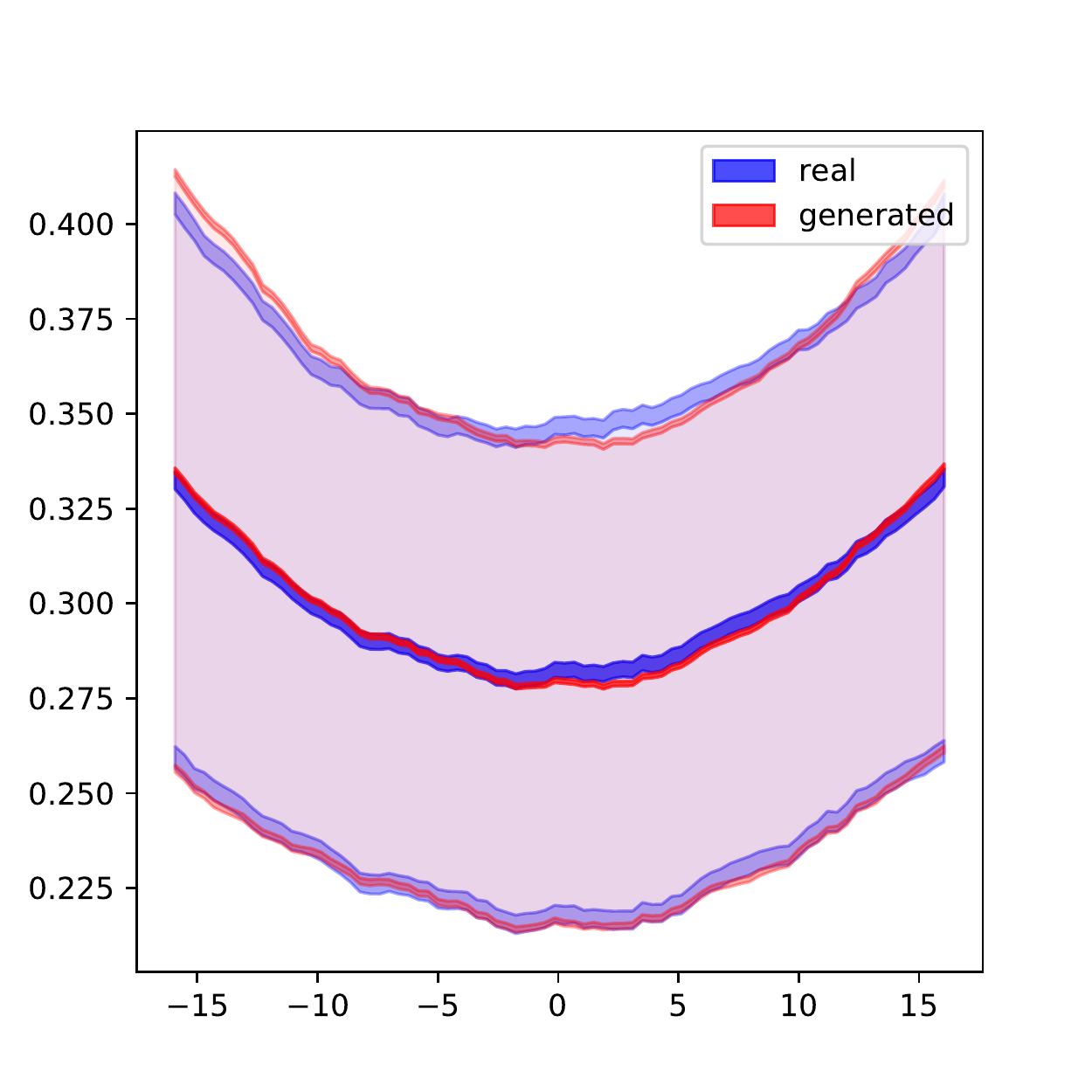}} &
\raisebox{-.5\height}{\includegraphics[width=.18\textwidth,trim=0 10 35 30,clip]{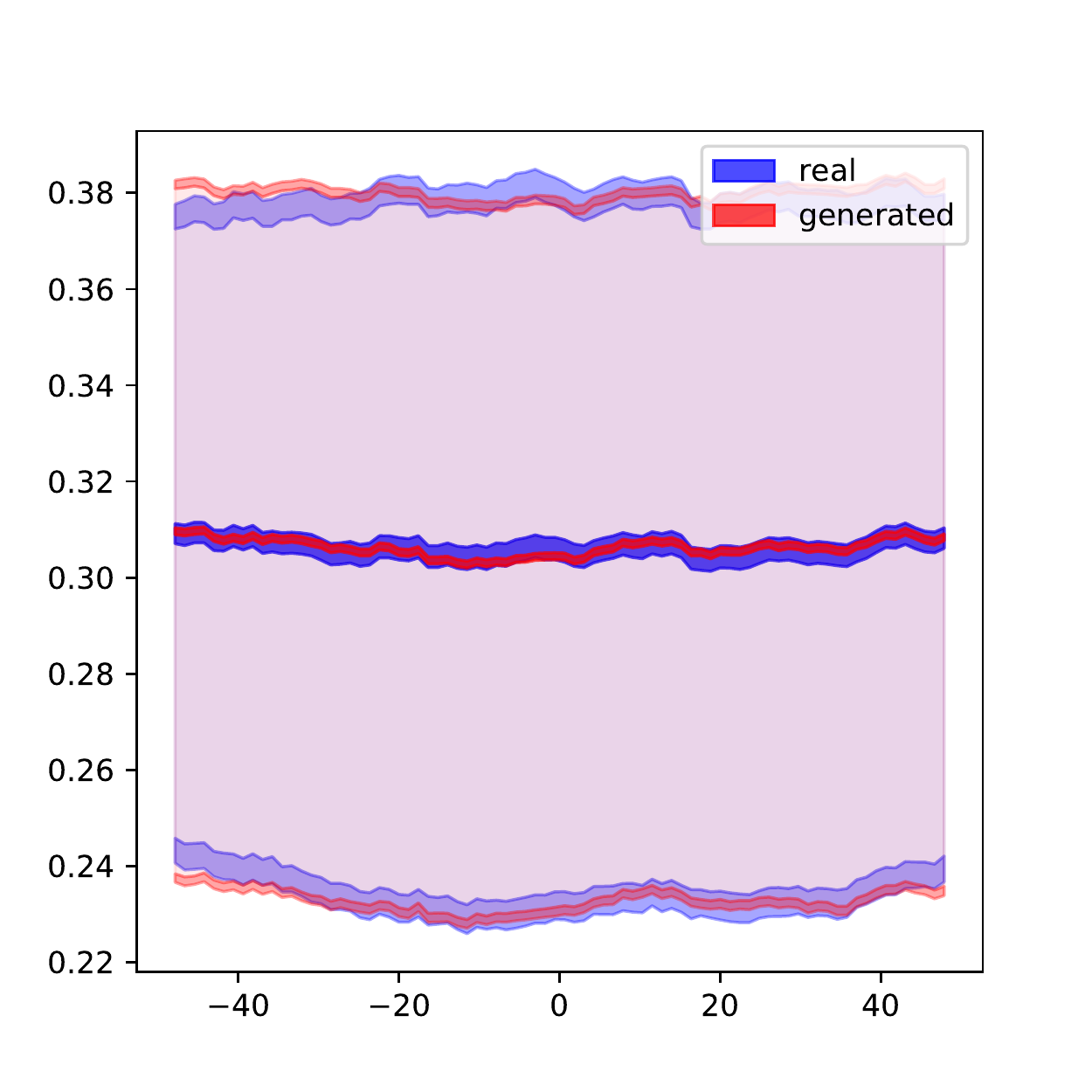}} &
\raisebox{-.5\height}{\includegraphics[width=.18\textwidth,trim=0 10 35 30,clip]{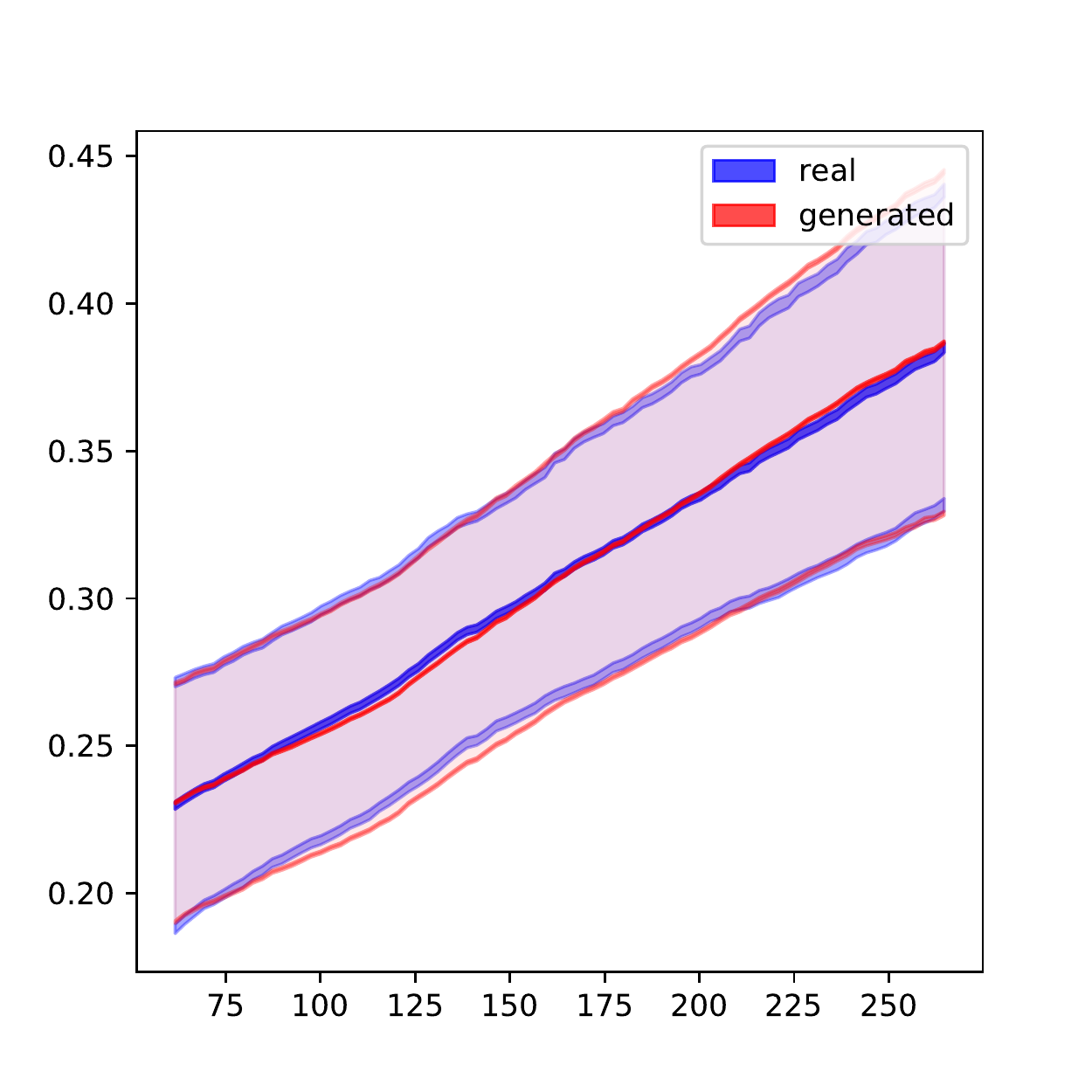}} &
\raisebox{-.5\height}{\includegraphics[width=.18\textwidth,trim=0 10 35 30,clip]{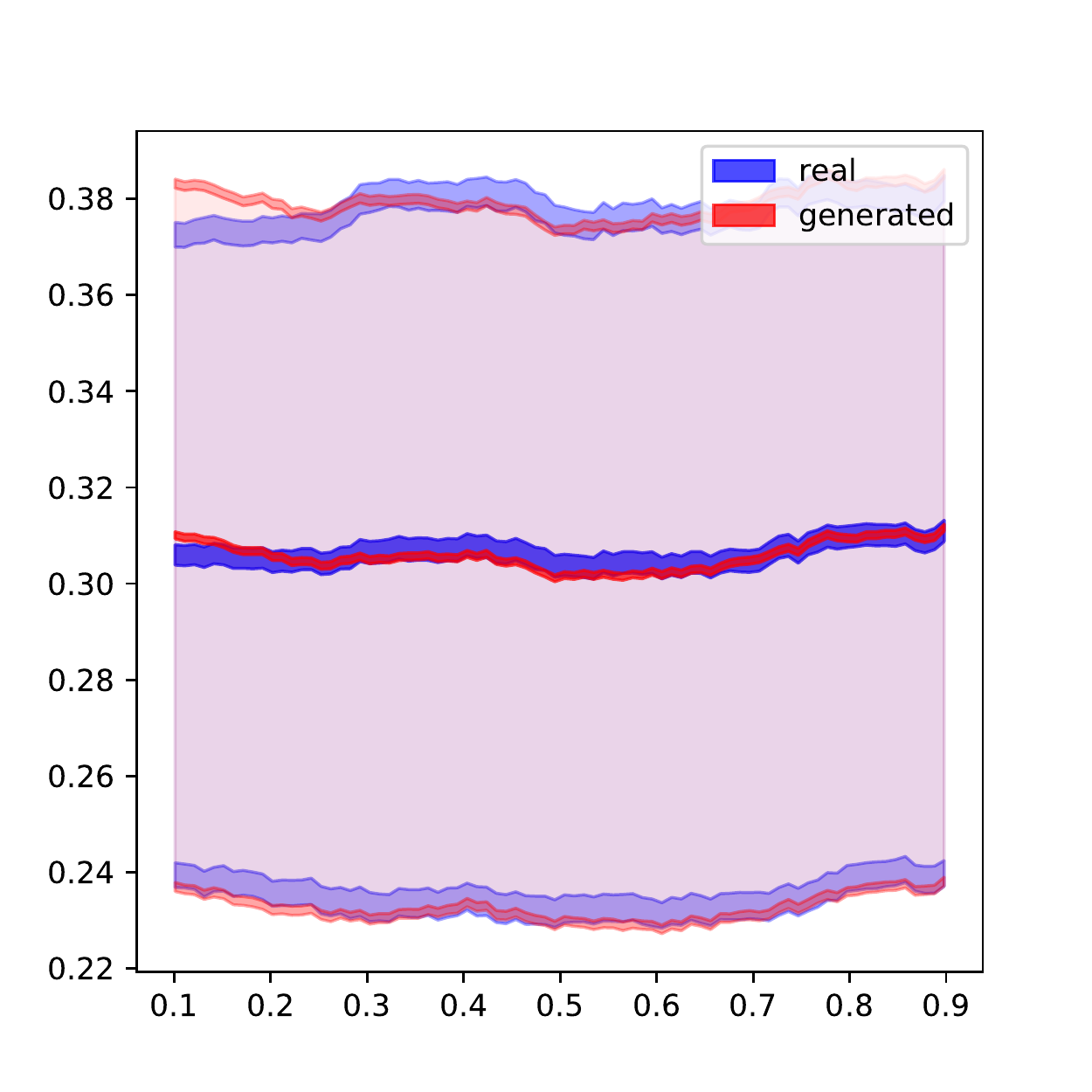}} &
\raisebox{-.5\height}{\includegraphics[width=.18\textwidth,trim=0 10 35 30,clip]{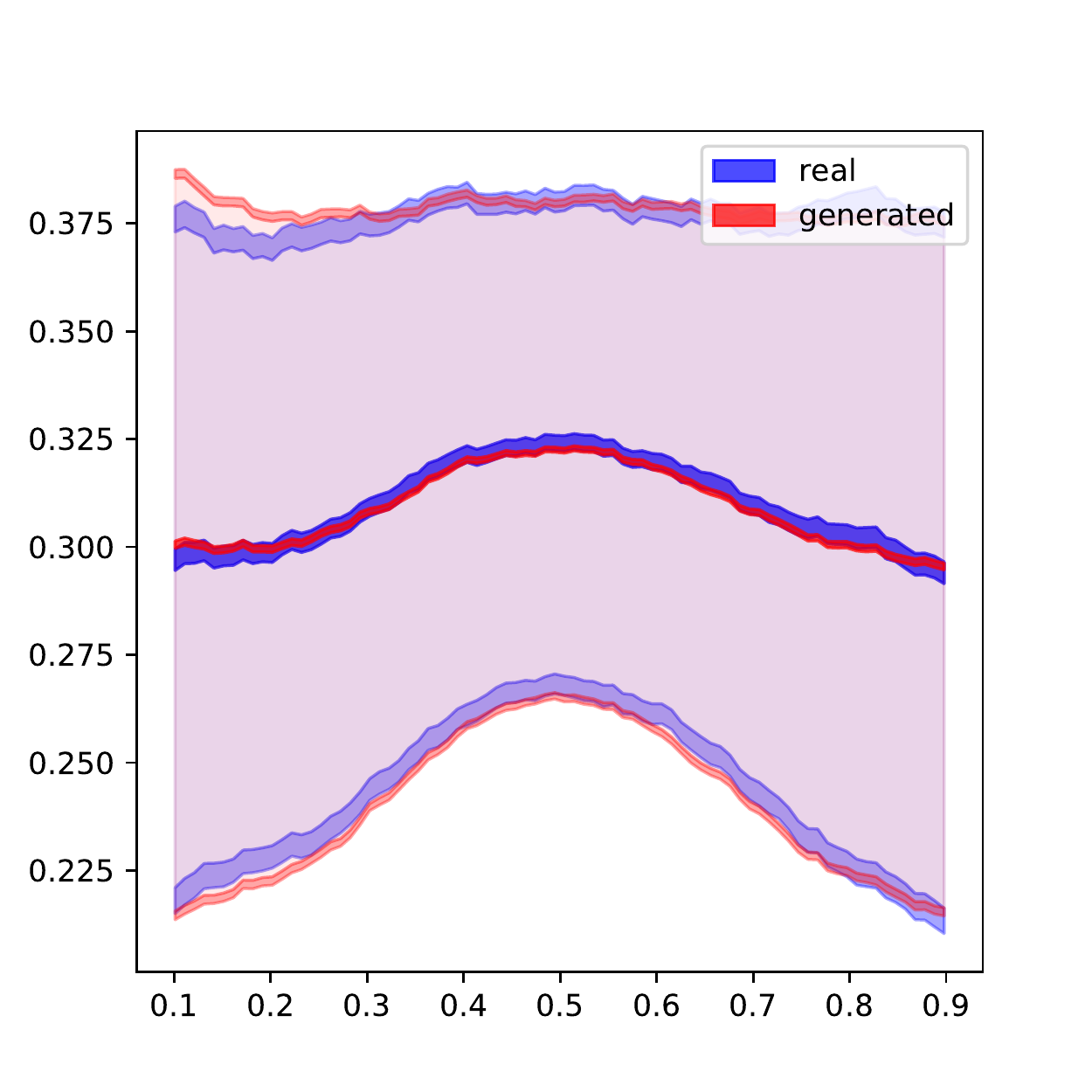}} \\ {\rotatebox[origin=c]{90}{\parbox{.18\textwidth}{\centering Sq. Time Width}}} &
\raisebox{-.5\height}{\includegraphics[width=.18\textwidth,trim=0 10 35 30,clip]{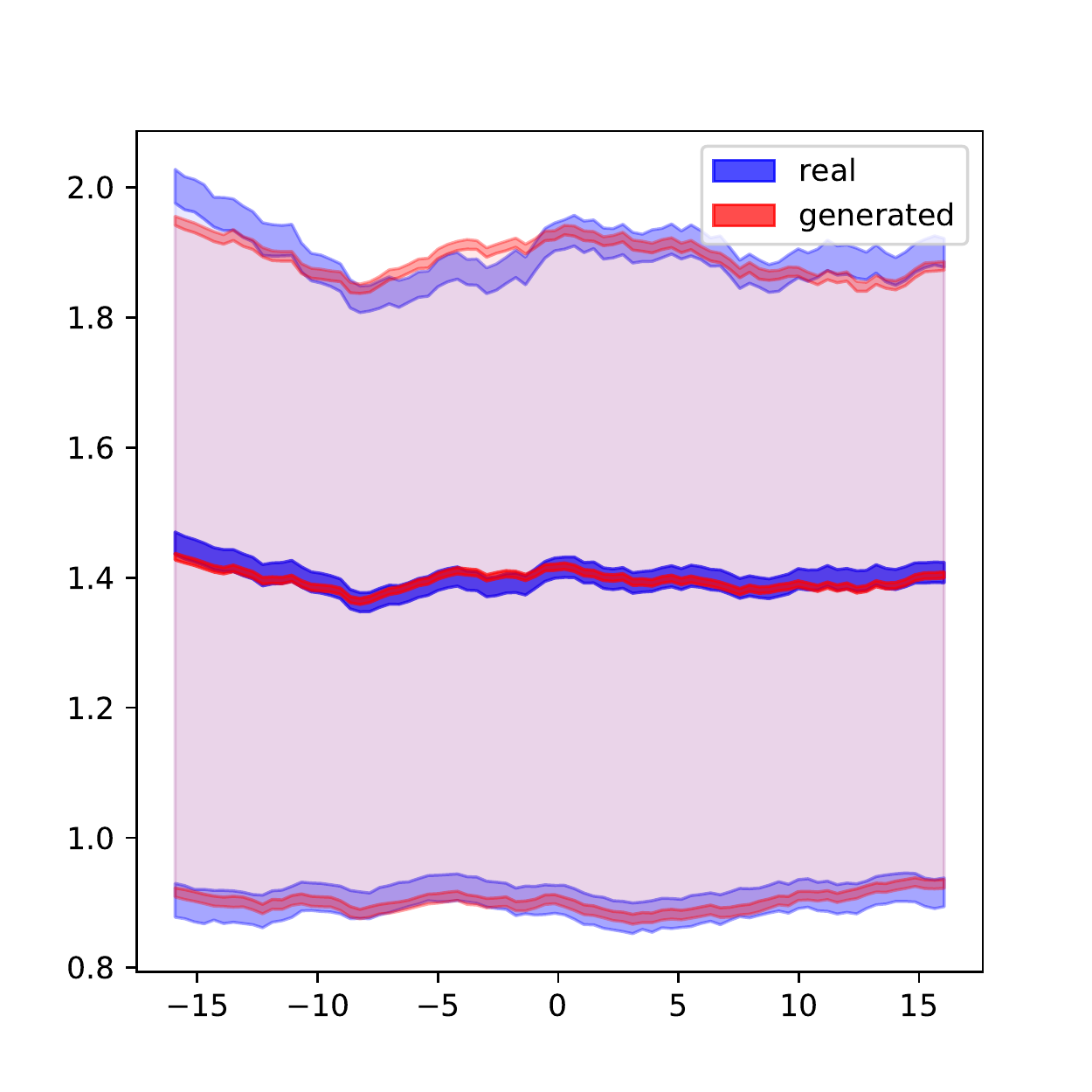}} &
\raisebox{-.5\height}{\includegraphics[width=.18\textwidth,trim=0 10 35 30,clip]{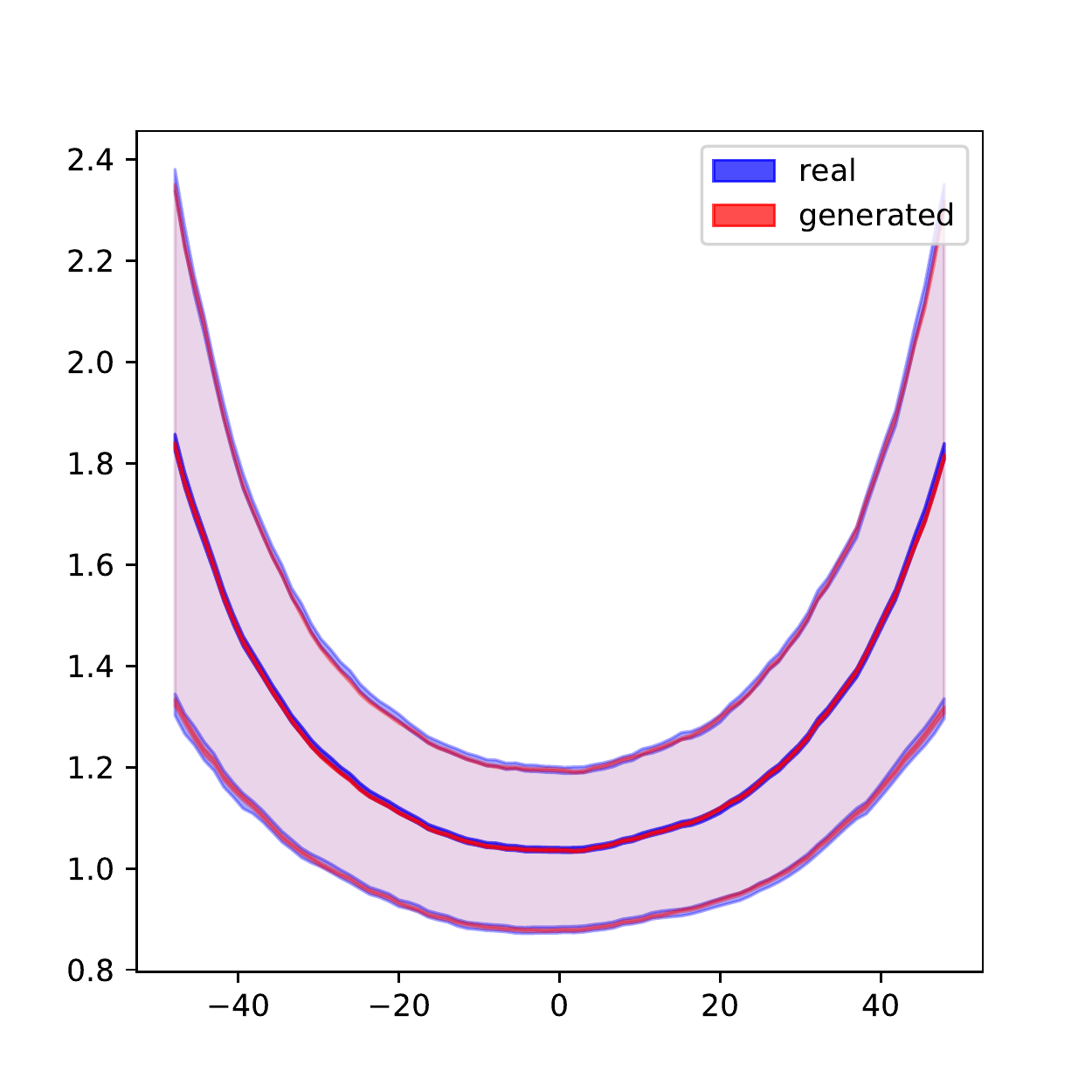}} &
\raisebox{-.5\height}{\includegraphics[width=.18\textwidth,trim=0 10 35 30,clip]{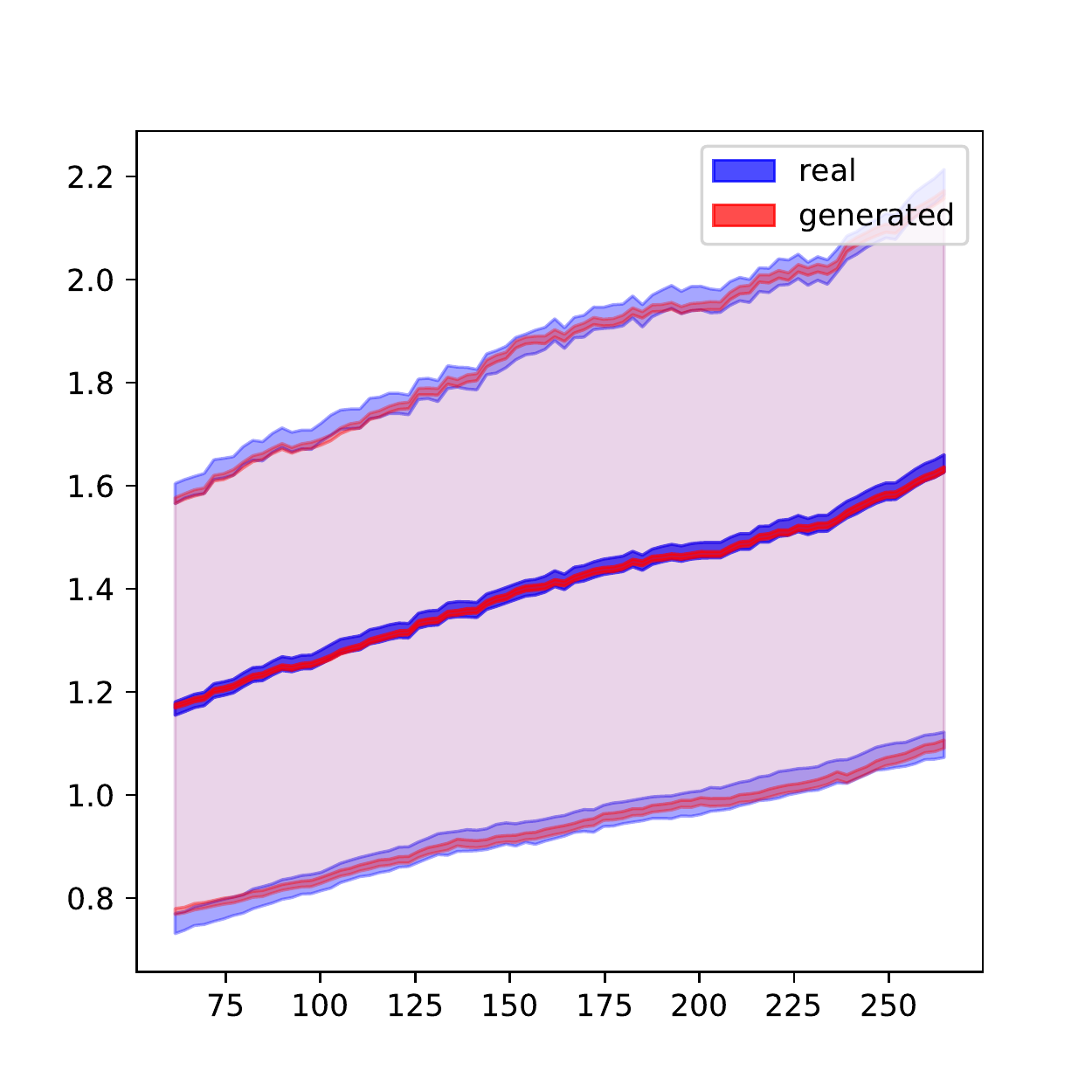}} &
\raisebox{-.5\height}{\includegraphics[width=.18\textwidth,trim=0 10 35 30,clip]{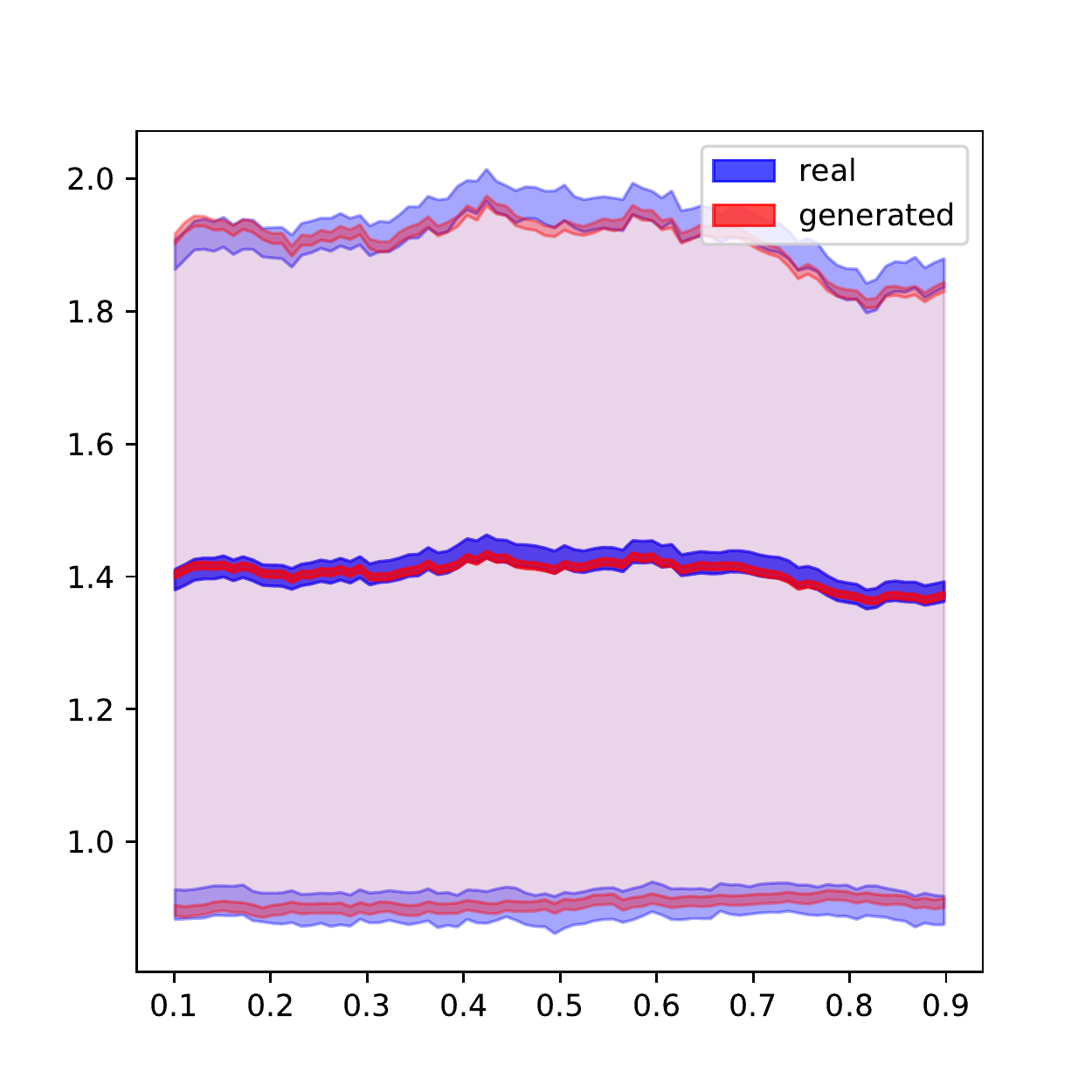}} &
\raisebox{-.5\height}{\includegraphics[width=.18\textwidth,trim=0 10 35 30,clip]{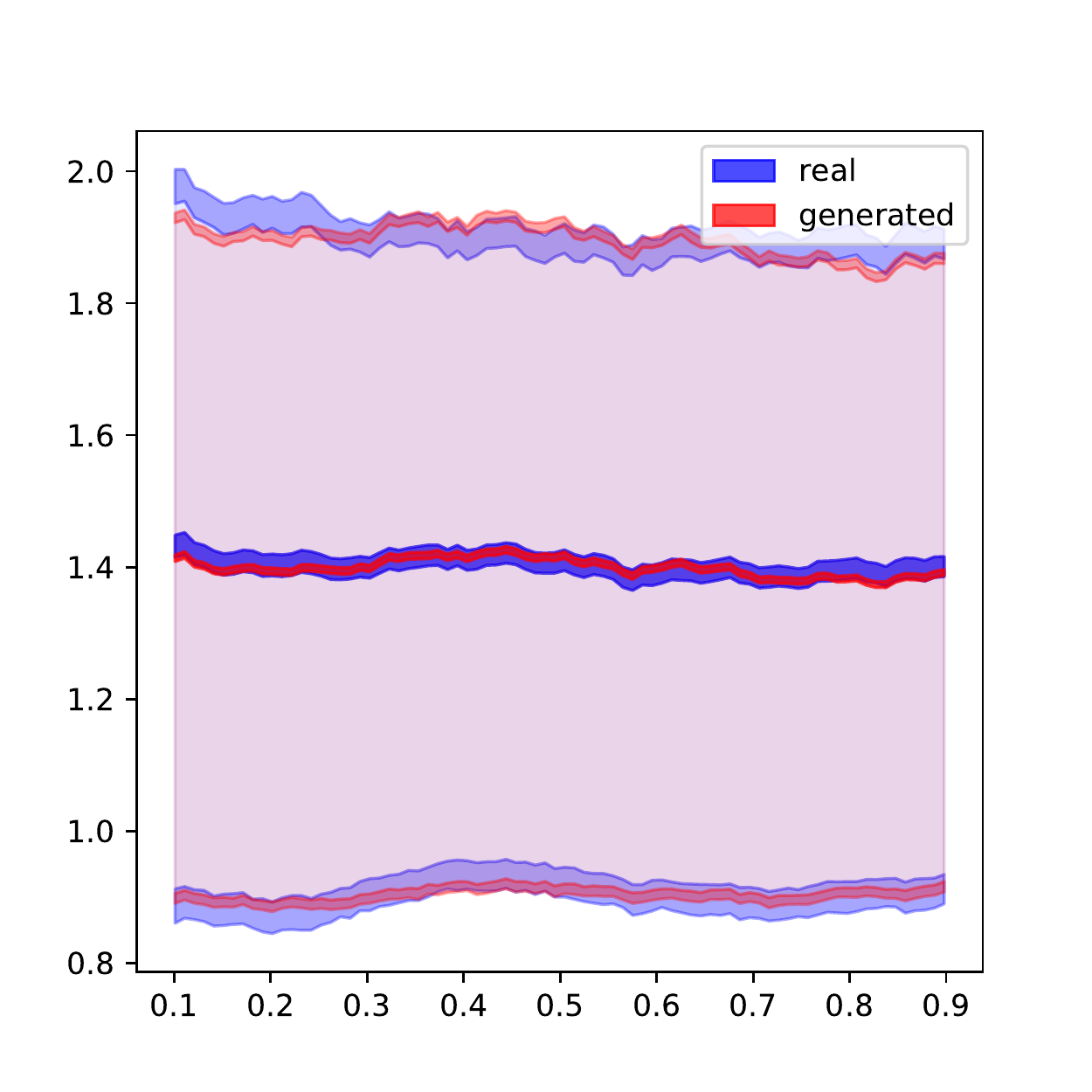}} \\ {\rotatebox[origin=c]{90}{\parbox{.18\textwidth}{\centering Pad-Time Covariance}}} &
\raisebox{-.5\height}{\includegraphics[width=.18\textwidth,trim=0 10 35 30,clip]{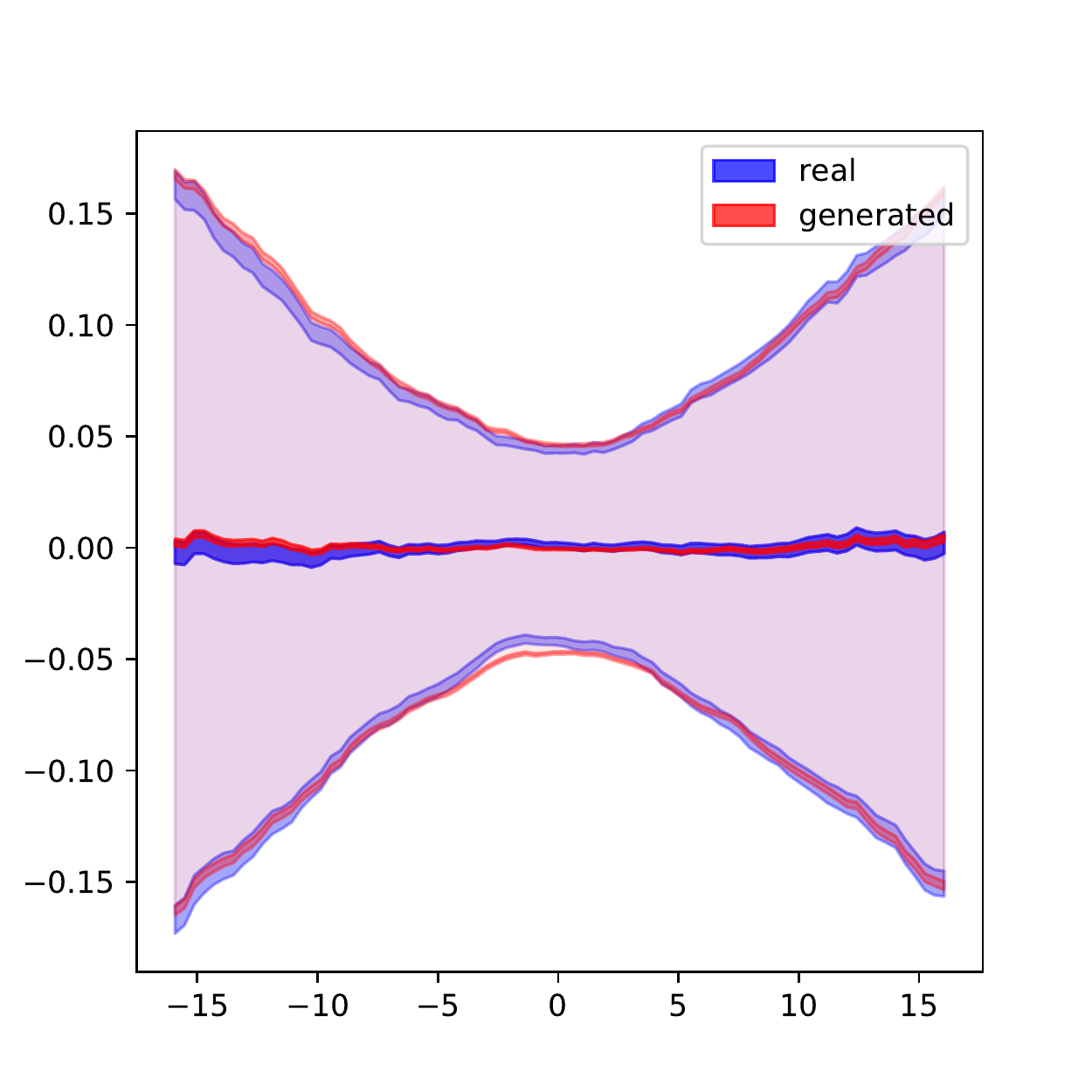}} &
\raisebox{-.5\height}{\includegraphics[width=.18\textwidth,trim=0 10 35 30,clip]{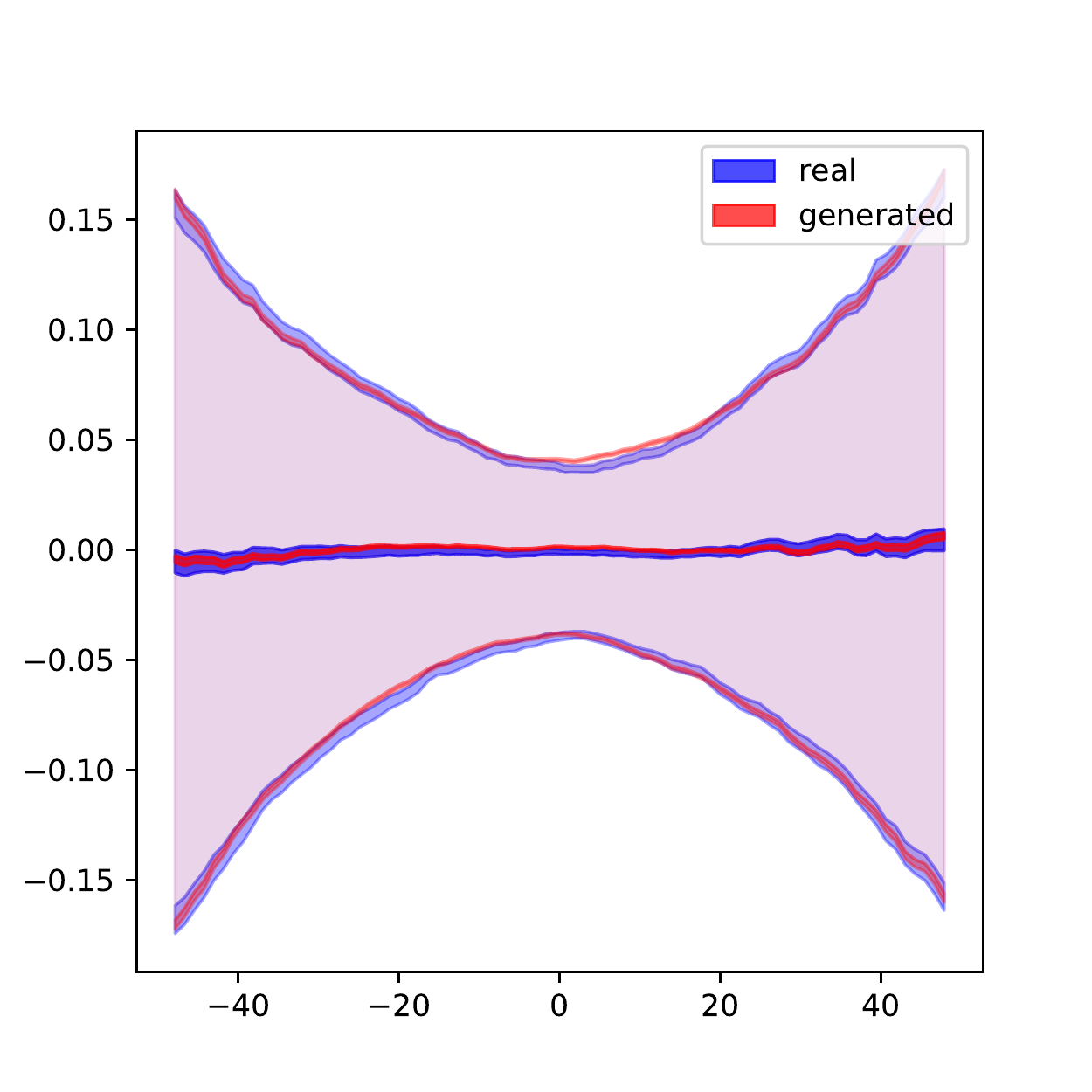}} &
\raisebox{-.5\height}{\includegraphics[width=.18\textwidth,trim=0 10 35 30,clip]{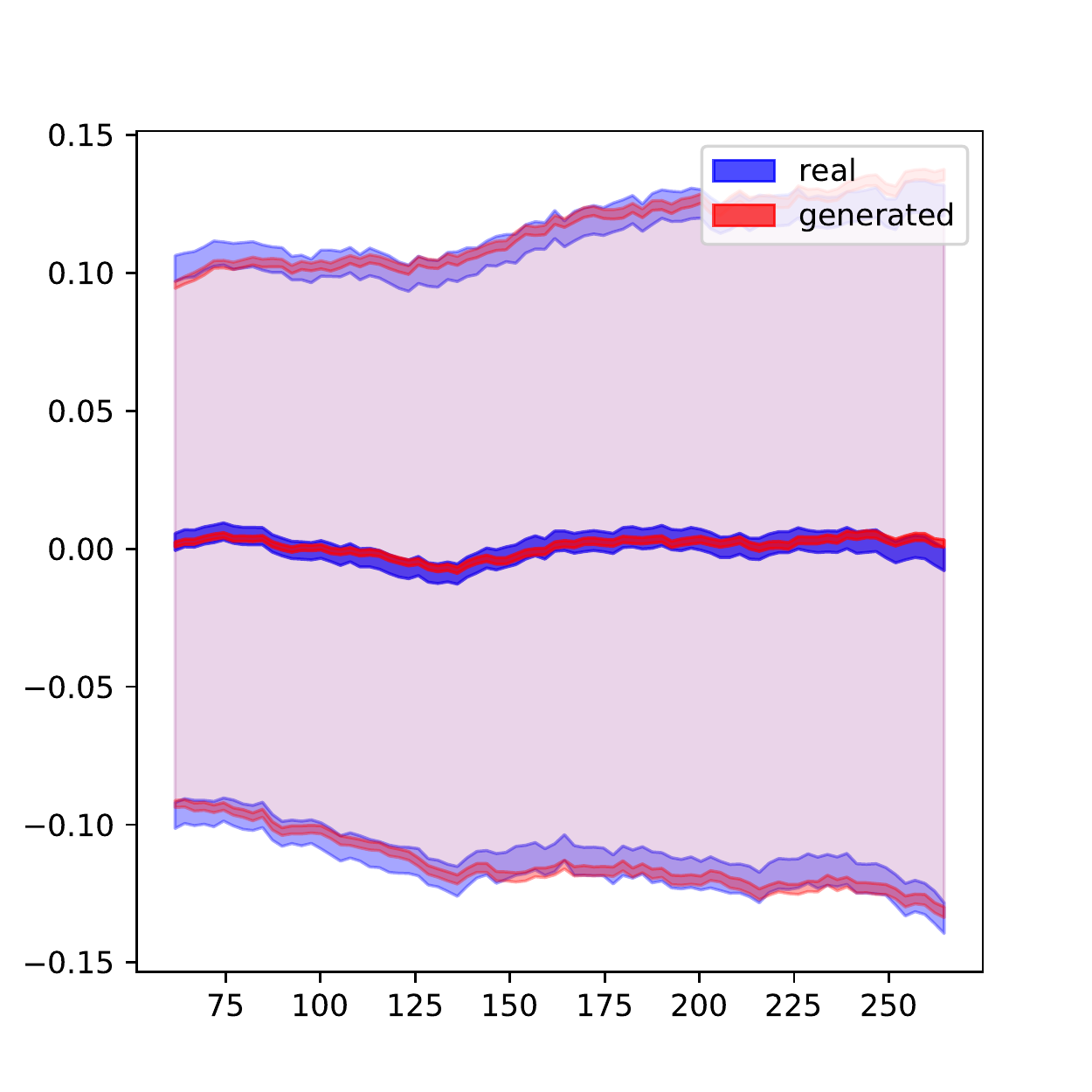}} &
\raisebox{-.5\height}{\includegraphics[width=.18\textwidth,trim=0 10 35 30,clip]{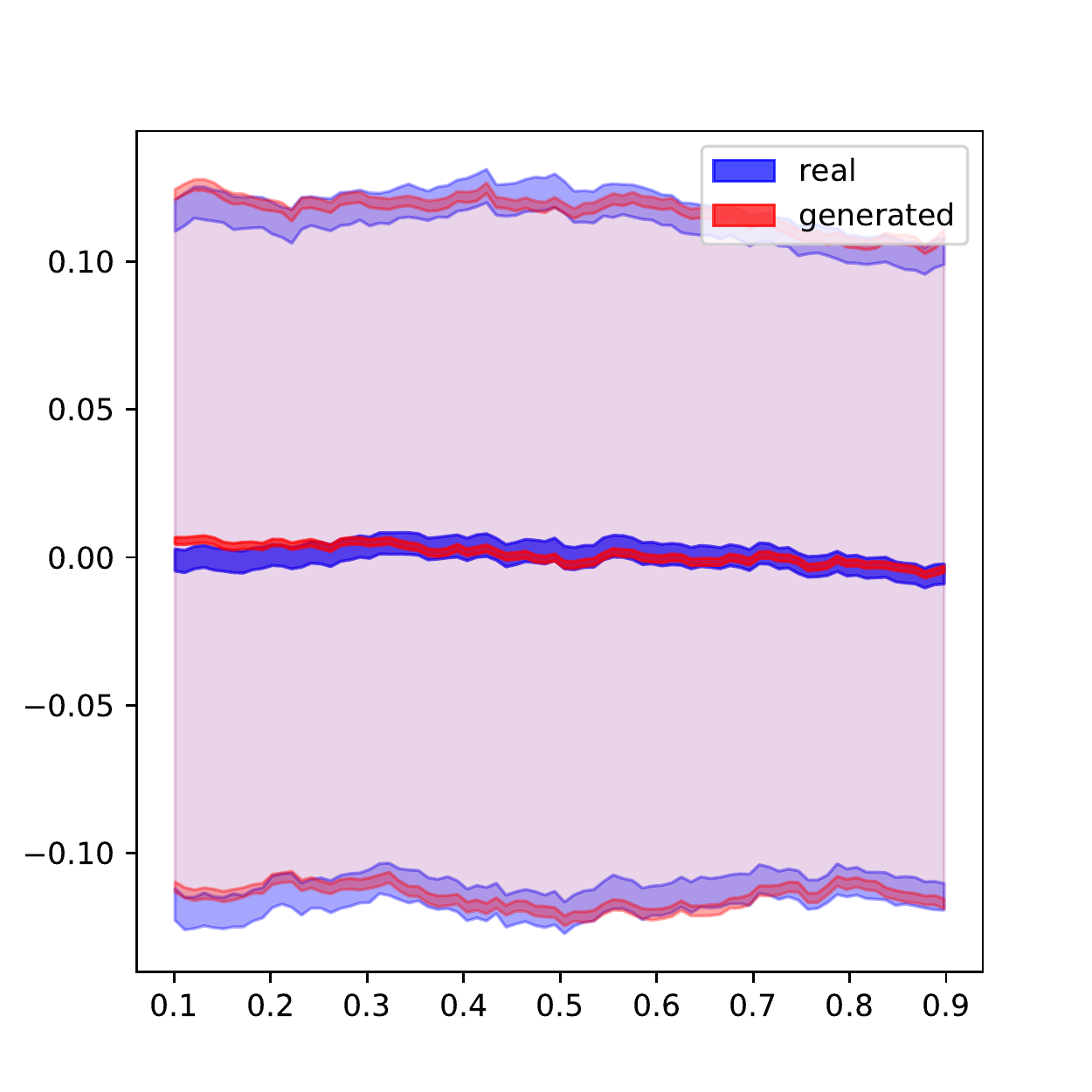}} &
\raisebox{-.5\height}{\includegraphics[width=.18\textwidth,trim=0 10 35 30,clip]{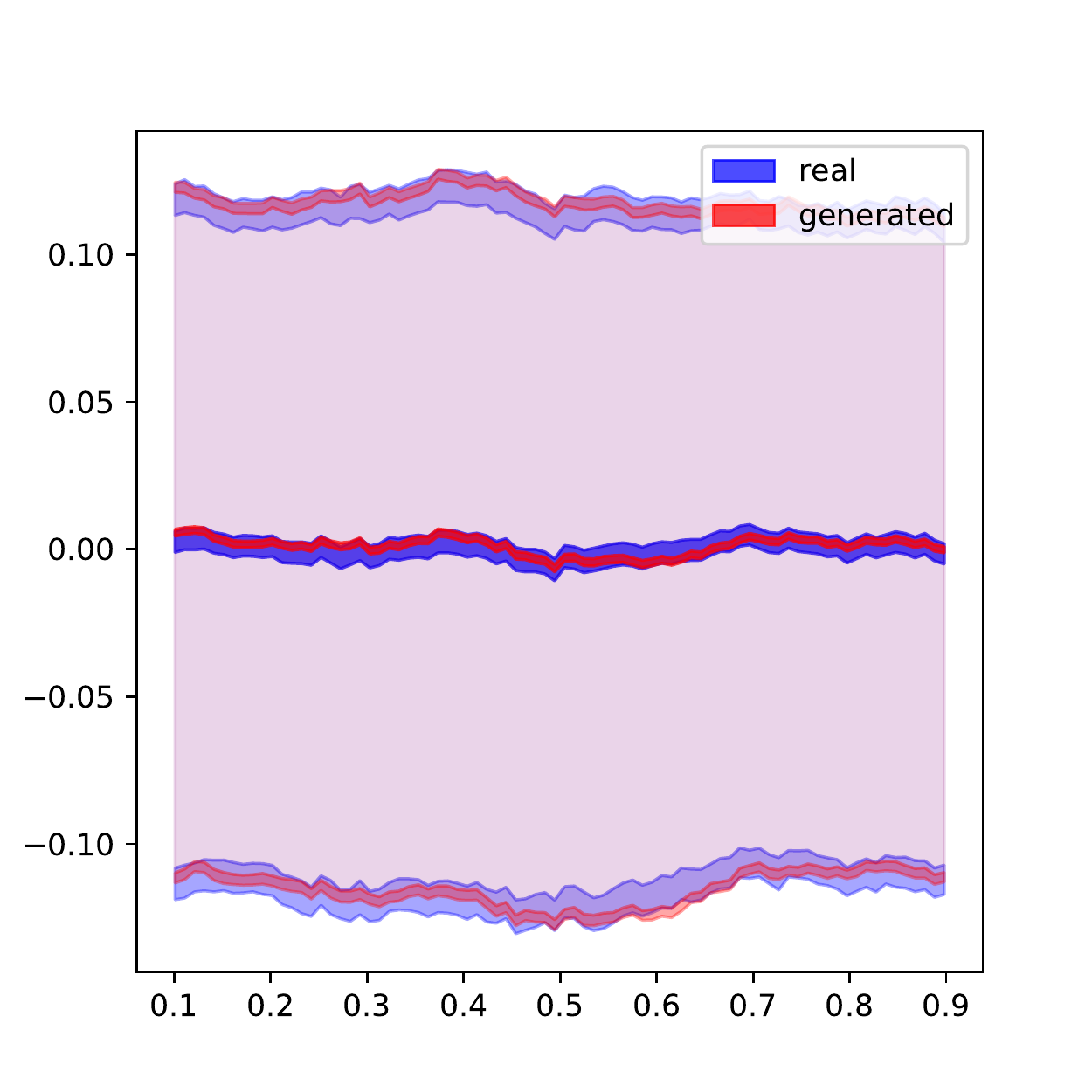}} \\ {\rotatebox[origin=c]{90}{\parbox{.18\textwidth}{\centering Integrated amplitude}}} &
\raisebox{-.5\height}{\includegraphics[width=.18\textwidth,trim=0 10 35 30,clip]{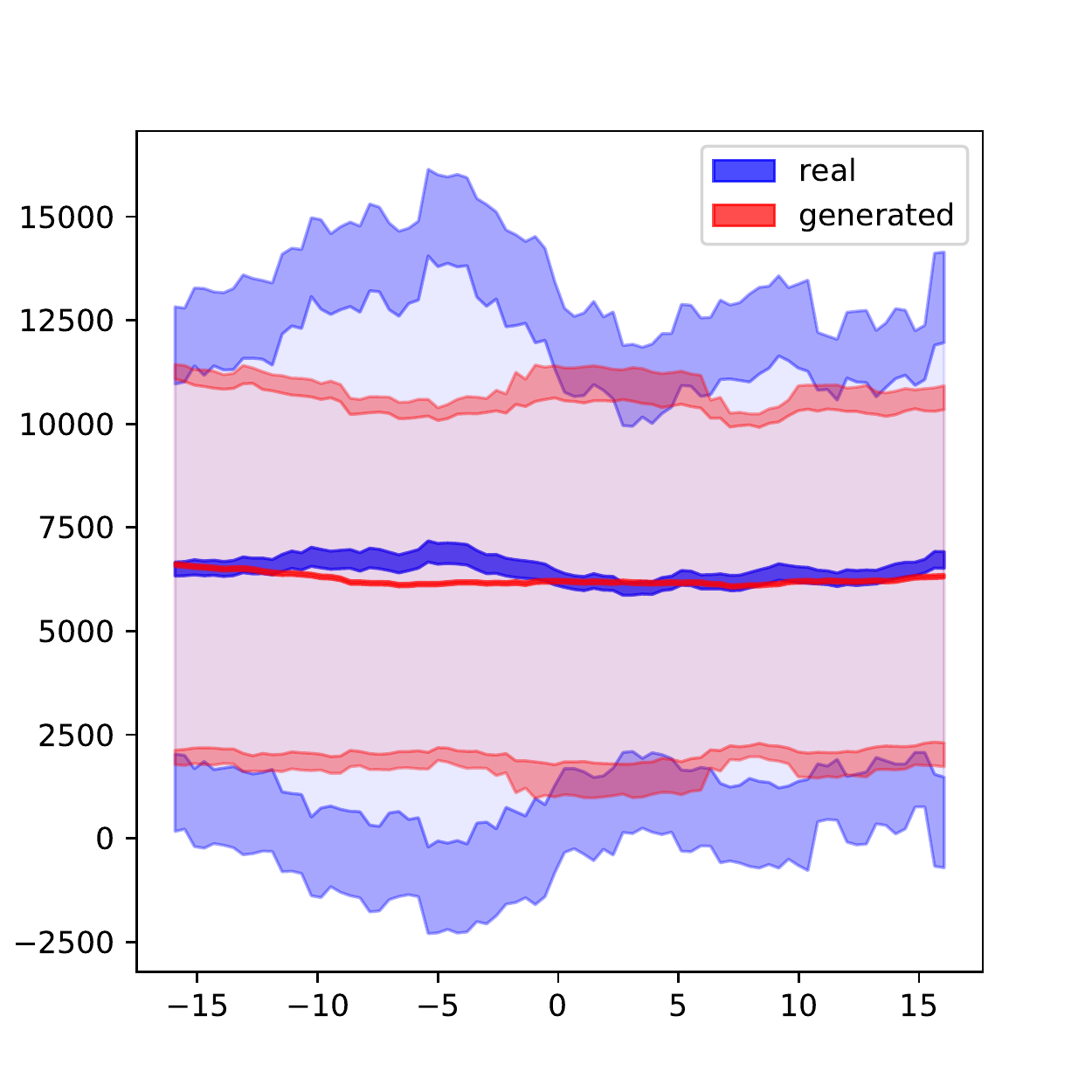}} &
\raisebox{-.5\height}{\includegraphics[width=.18\textwidth,trim=0 10 35 30,clip]{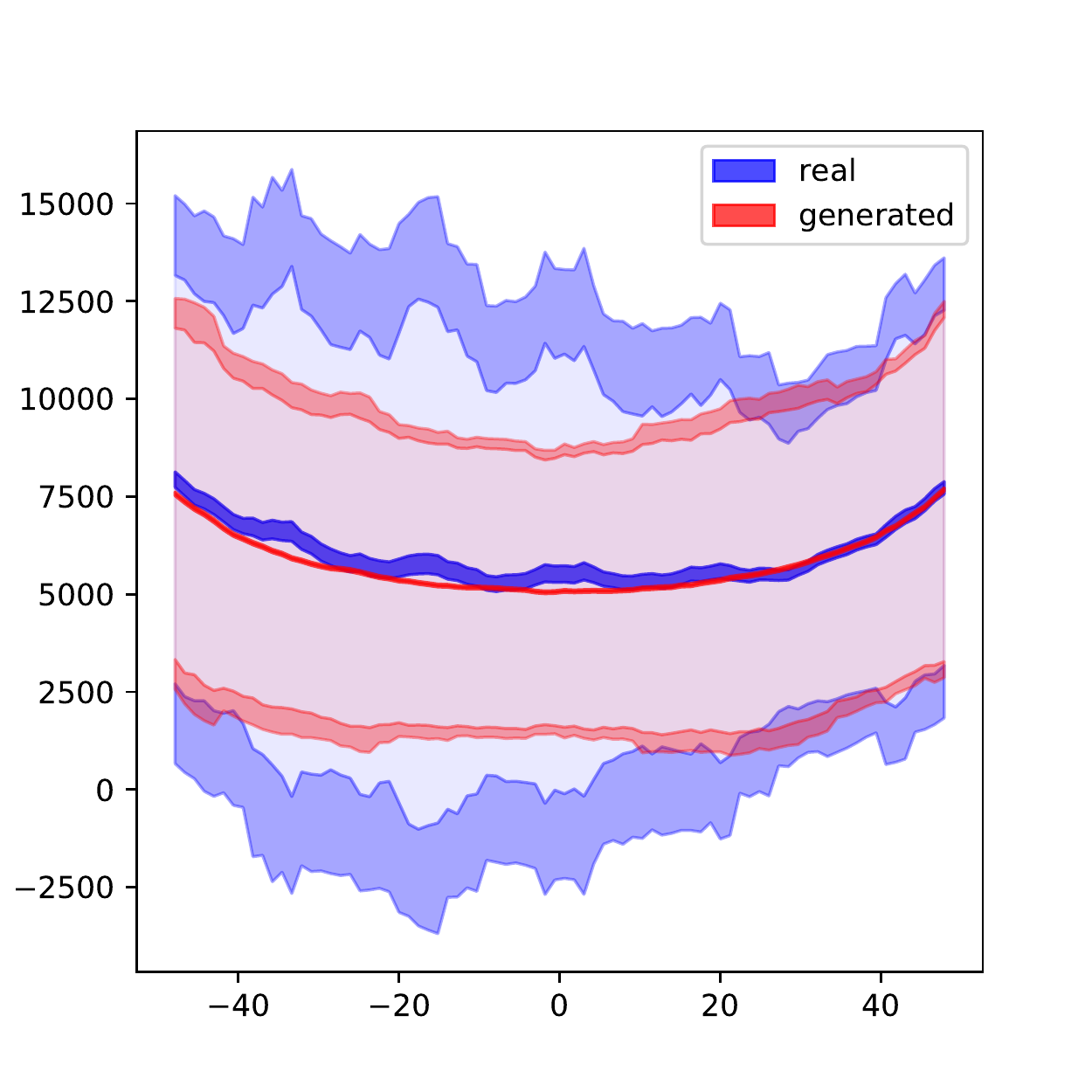}} &
\raisebox{-.5\height}{\includegraphics[width=.18\textwidth,trim=0 10 35 30,clip]{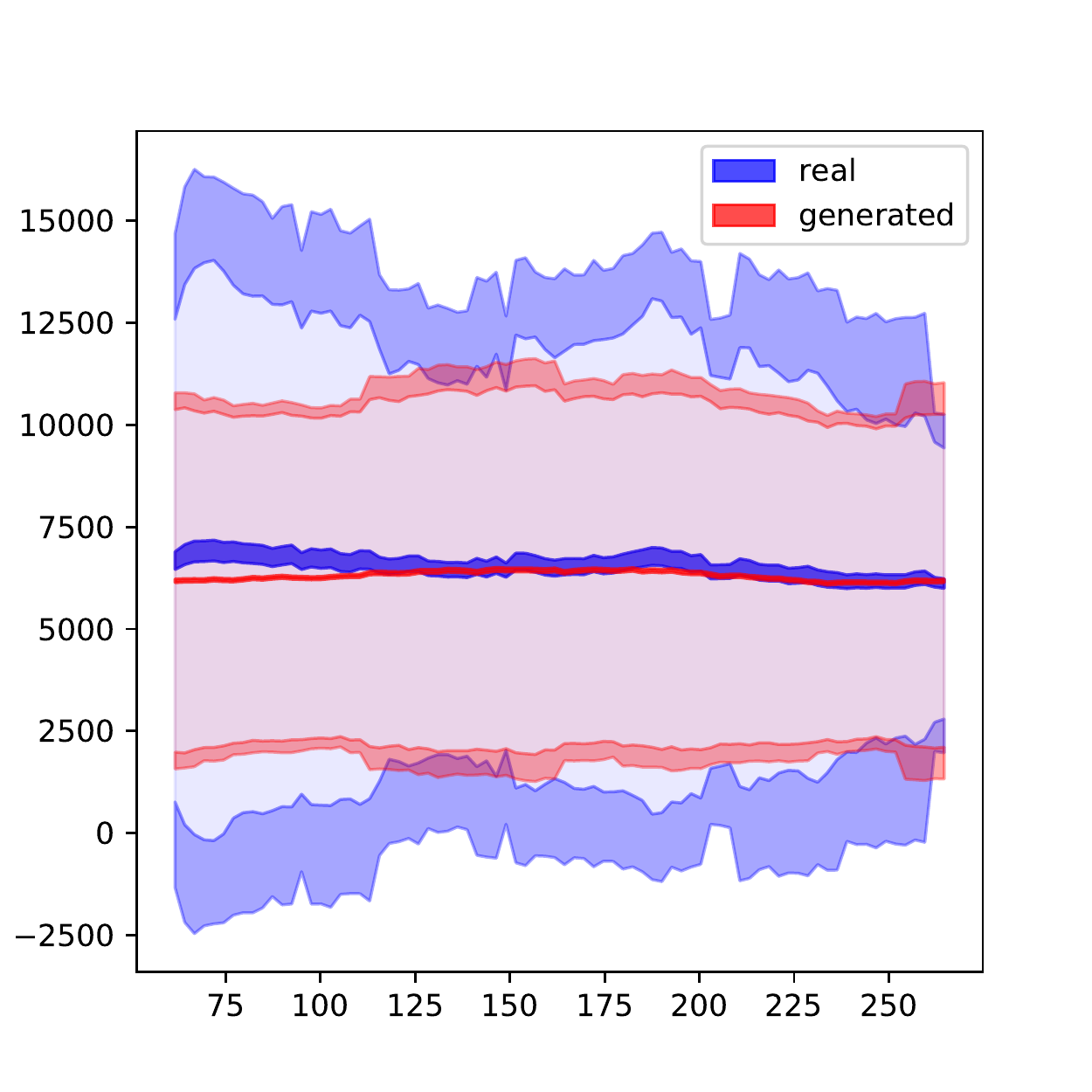}} &
\raisebox{-.5\height}{\includegraphics[width=.18\textwidth,trim=0 10 35 30,clip]{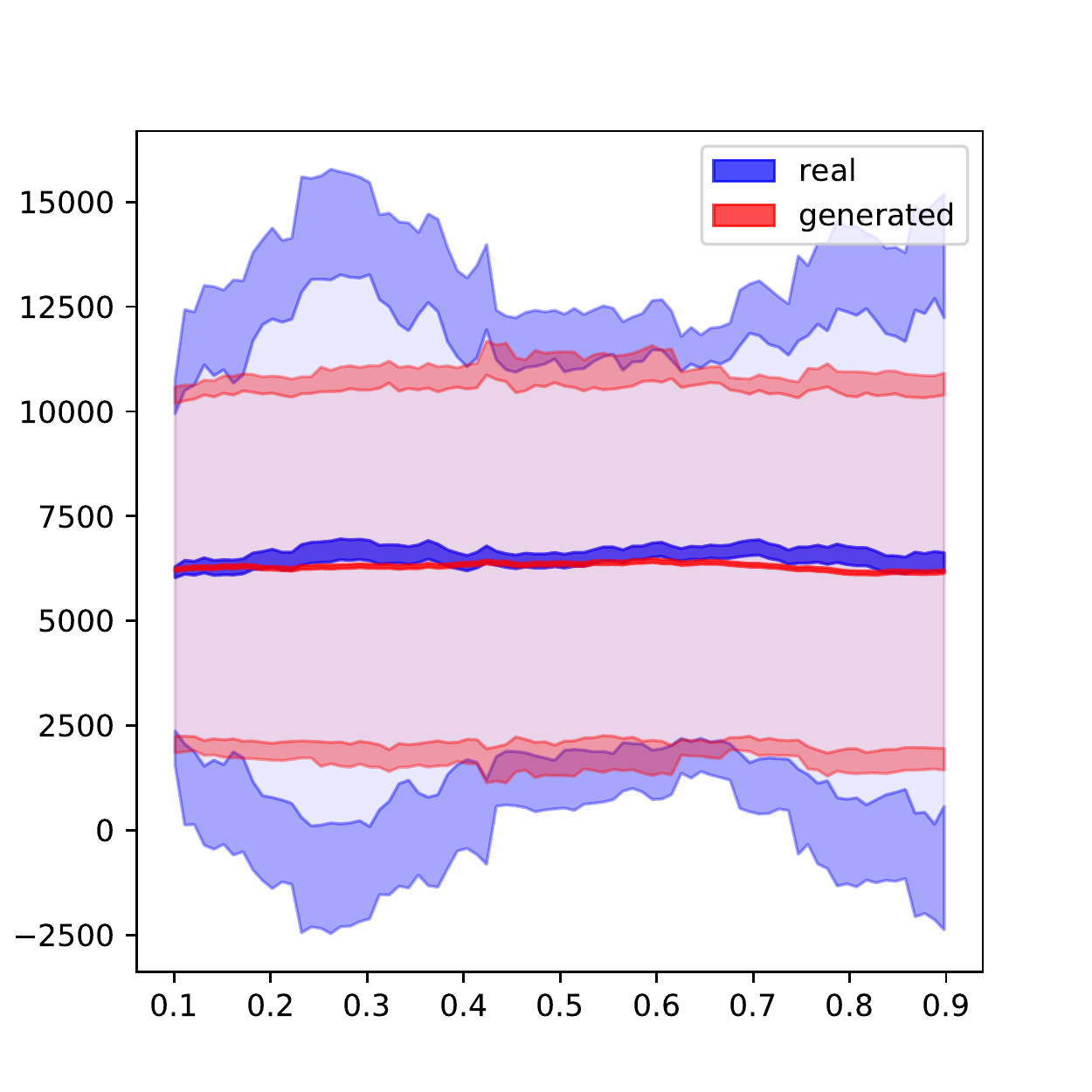}} &
\raisebox{-.5\height}{\includegraphics[width=.18\textwidth,trim=0 10 35 30,clip]{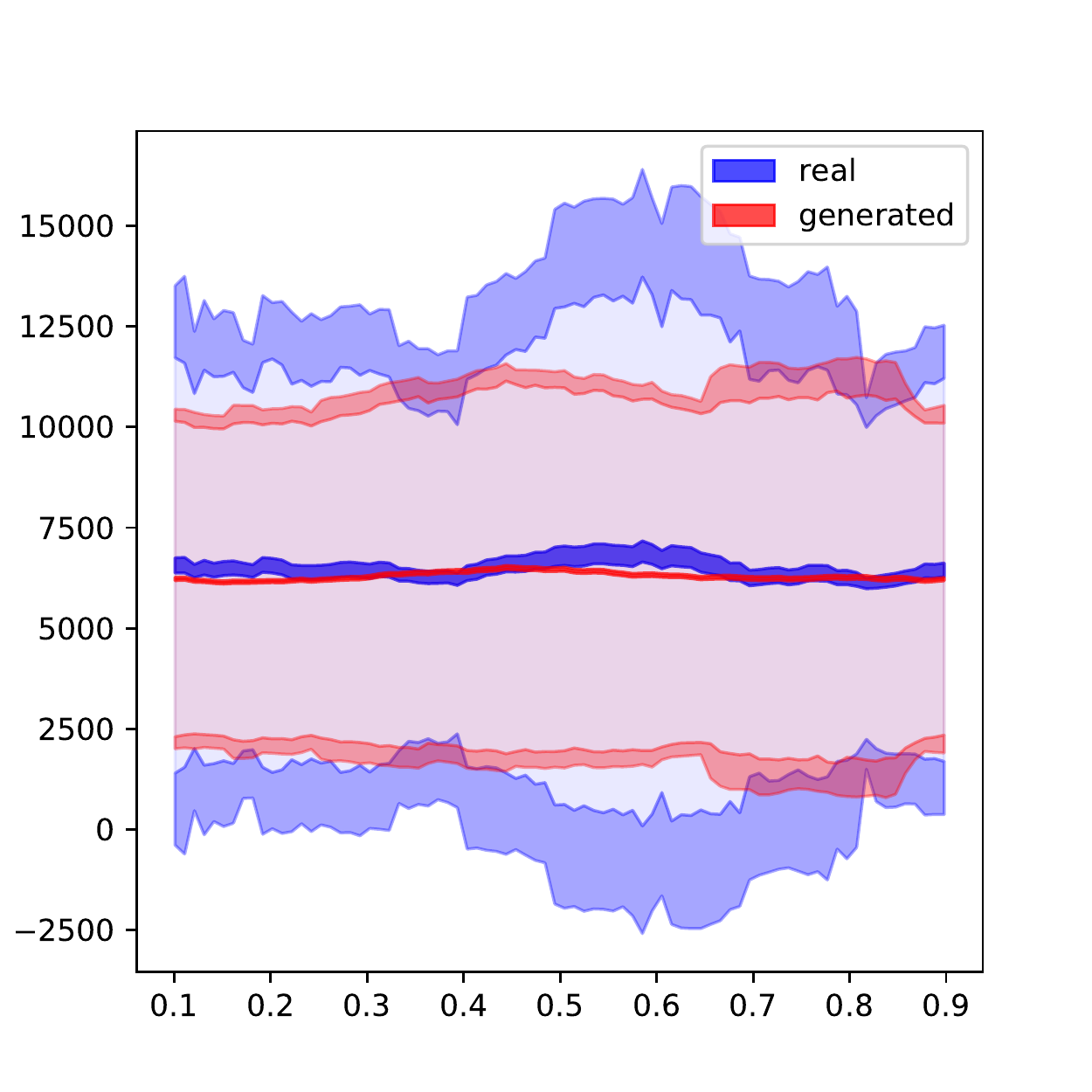}} \\
~ &
\hspace{10pt}\parbox{.15\textwidth}{\centering Crossing angle [deg]} &
\hspace{10pt}\parbox{.15\textwidth}{\centering Dip angle [deg]} &
\hspace{10pt}\parbox{.15\textwidth}{\centering Drift length [time bins]} &
\hspace{10pt}\parbox{.15\textwidth}{\centering Drift length fraction} &
\hspace{10pt}\parbox{.15\textwidth}{\centering Pad coordinate fraction}
\end{tabular}
\endgroup
\caption{\label{fig:preevalplots} Profiles of the validation metric distributions as the function of the input variables. For each metric $\xi$, we show its average value $\mu_\xi$ (middle thick line), as well as the average shifted up and down by a standard deviation of the metric distribution $\mu_\xi\pm\sigma_\xi$ (top and bottom thick lines). Line thickness denotes the statistical uncertainty of the corresponding value. For a smoother representation, the values were calculated in overlapping running windows over the input variables (100 bins total, 20-bin window size).}
\end{figure*}

To make a more precise quantitative evaluation, we introduce a set of metrics that we profile as a function of input variables and compare between the generated and validation data. For each pad response image, we calculate the 1st order moments, i.e. the pad and time coordinates of the response distribution barycenter, the 2nd order moments, i.e. the squared widths of the response distribution and covariance between the pad and time coordinates, and the integrated amplitude. The profiles of these quantities can be seen in Fig.~\ref{fig:preevalplots}.

Overall, the plots in Fig.~\ref{fig:preevalplots} demonstrate a good agreement between the averages of the metric distributions, and also a reasonable agreement between their widths. The pad response barycenters and widths along pad row and time directions are well reproduced, which should translate to the accurate simulation of the coordinate resolution and two-track resolution of the TPC. The most notable biases may be found in the variances of the integrated amplitude distributions, as the amplitudes span several orders of magnitude and hence are particularly difficult to model precisely. Biases in the amplitude modeling can affect the $dE/dx$ measurements, and thus the model estimates should be used with care. It should be noted that our approach is aimed to replace the expensive simulation of the electron drift and electronics response modeling, rather than track propagation. This means that Geant3 energy deposits in the detector material can be used to scale the predicted integrated amplitude and therefore account for any mismodeling of the amplitude introduced by the GAN. This is the preferred solution that is used in the tests described below.

In order to evaluate the quality of our model with reconstructed detector objects, our model is integrated into the MPD software stack. We use our model to simulate TPC pad responses for tracks from central Au+Au collisions at $\sqrt{s_{NN}}=9$\,GeV generated with UrQMD~\cite{Bleicher:1999xi}. The neural network is used to simulate pad responses for the charged particles of a wide kinematic range in both long and short pads of the TPC, even though it is only trained on the responses in short pads from pions with a fixed transverse momentum. Other MPD subsystems are simulated with the detailed simulation procedures. The simulated detector responses are then fed into the standard reconstruction and track finding algorithms~\cite{Gertsenberger:2016llj}. The comparison with the detailed simulation is then carried out for the reconstructed tracks with $\left|y\right|< 0.5$, which have at least 20 hits reconstructed in the TPC and originate from the primary pions in the event.

\begin{figure*}
\centering
\begin{subfigure}{0.33\textwidth}
\centering
\includegraphics[width=\textwidth]{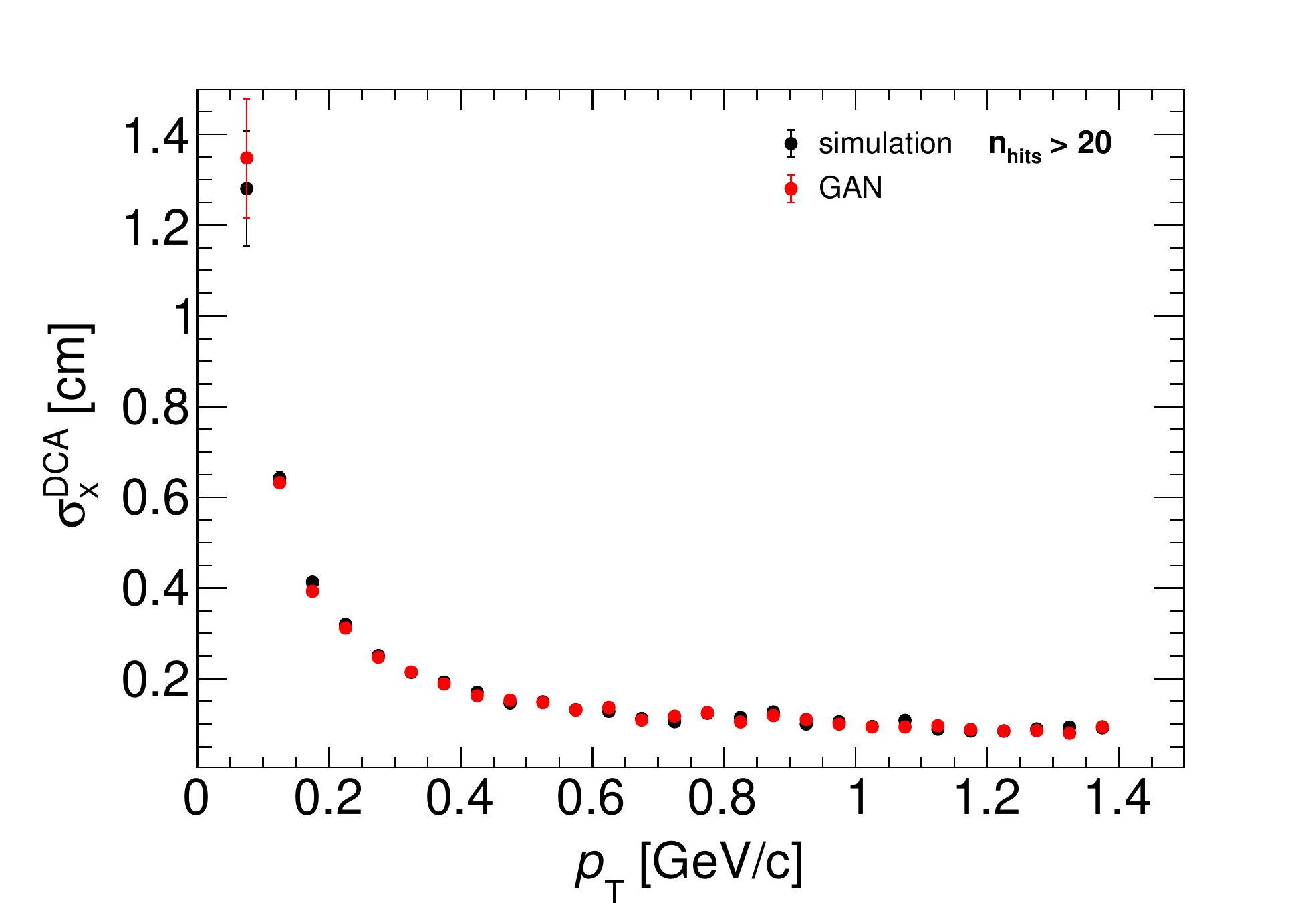}
\caption{Distance of closest approach resolution along $x$}
\label{fig:resolution:dcax}
\end{subfigure}
\begin{subfigure}{0.33\textwidth}
\centering
\includegraphics[width=\textwidth]{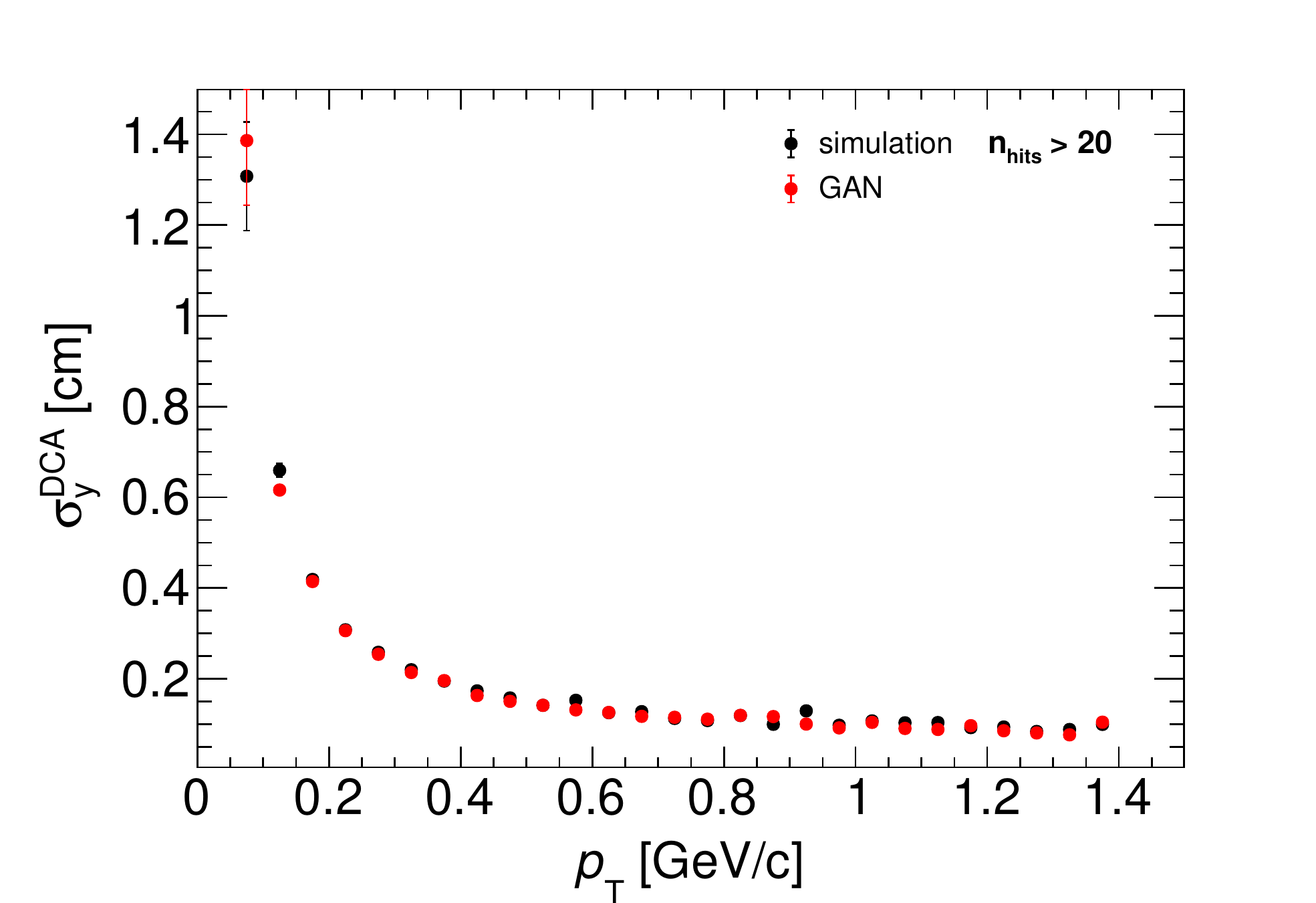}
\caption{Distance of closest approach resolution along $y$}
\label{fig:resolution:dcay}
\end{subfigure}
\begin{subfigure}{0.33\textwidth}
\centering
\includegraphics[width=\textwidth]{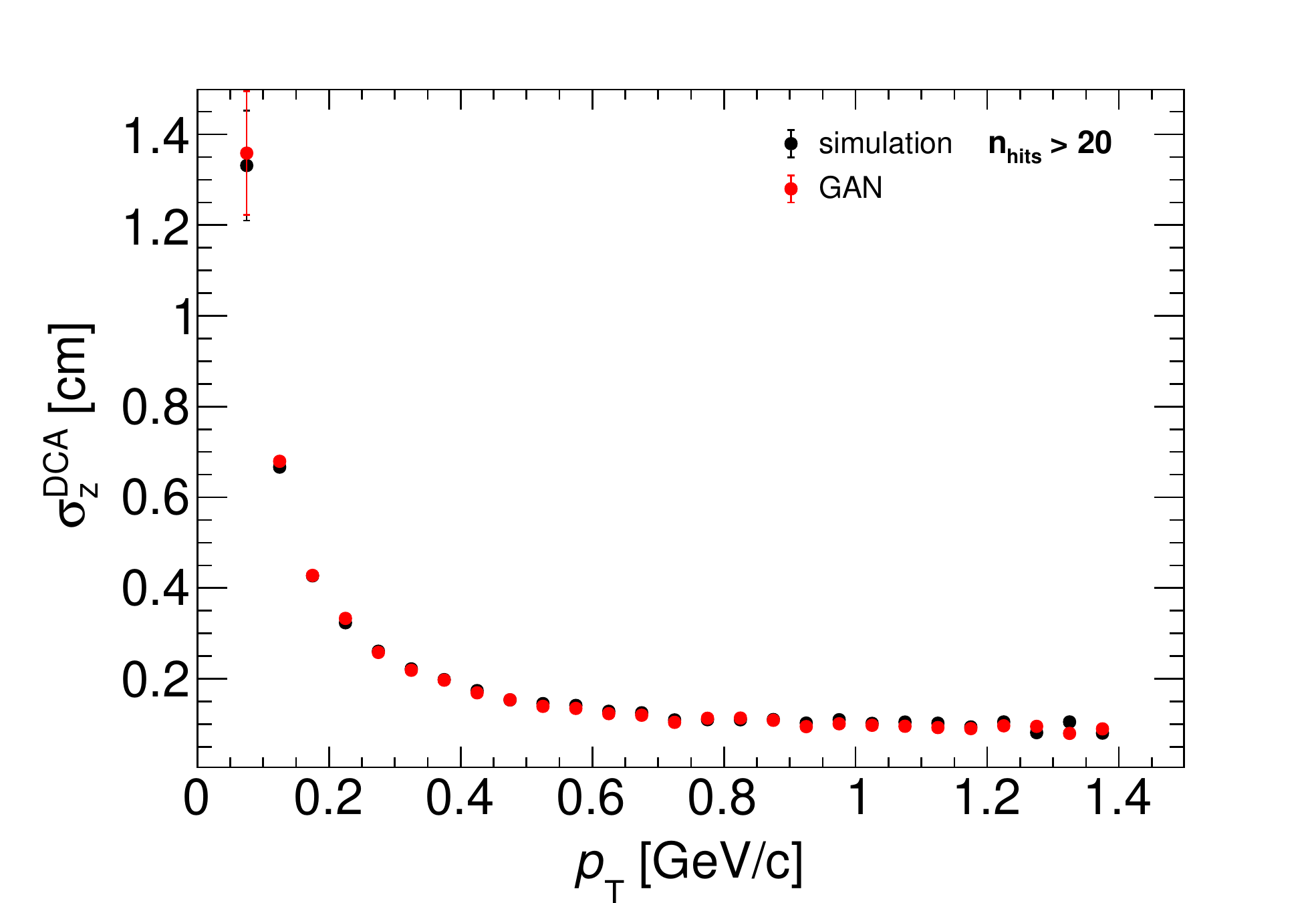}
\caption{Distance of closest approach resolution along $z$}
\label{fig:resolution:dcaz}
\end{subfigure}
\caption{\label{fig:resolution:dca} Distance of closest approach resolution as a function of the transverse momentum}
\end{figure*}

\begin{figure}
\begin{minipage}{\columnwidth}

\centering
\includegraphics[width=\textwidth]{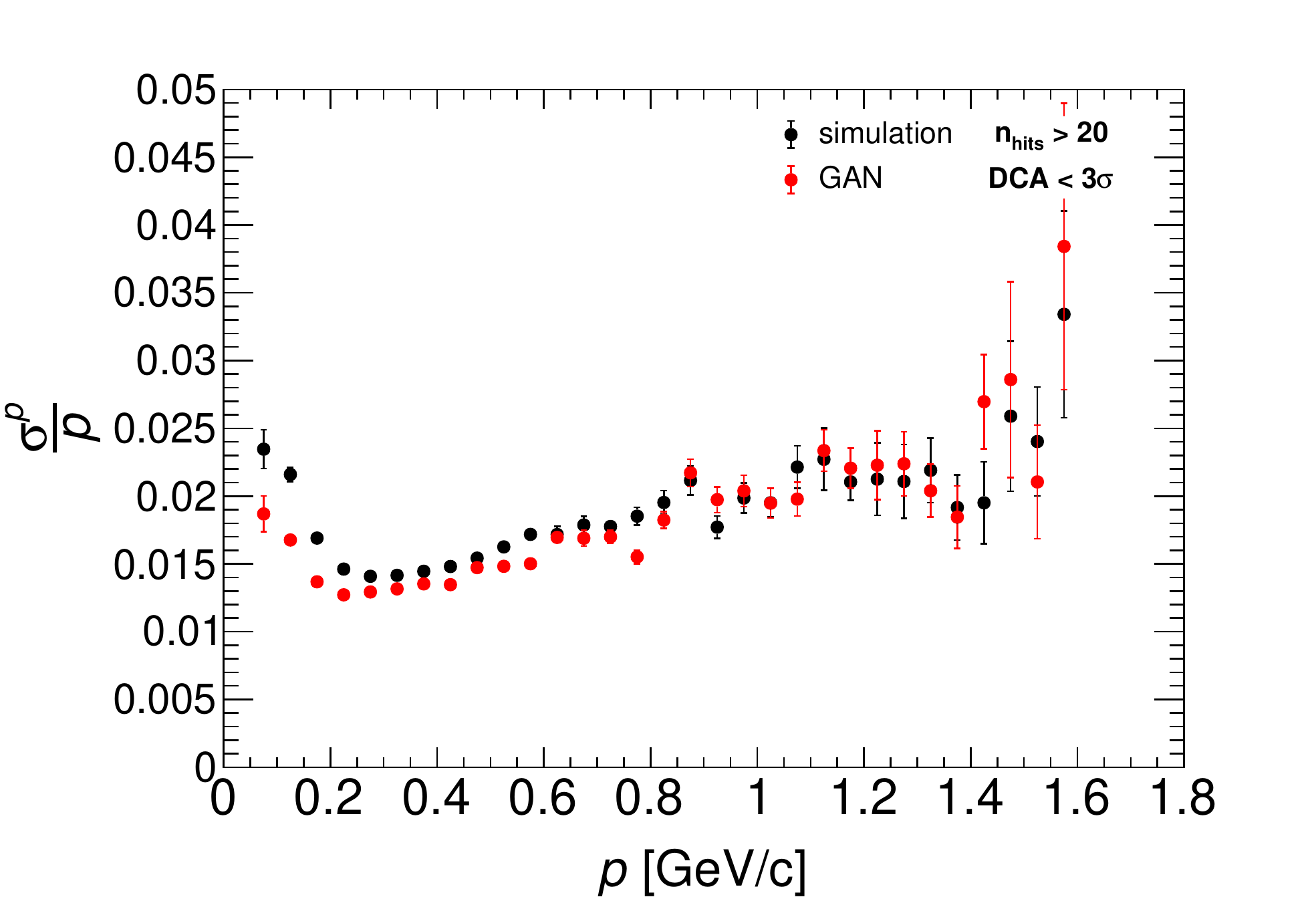}
\end{minipage}
\caption{\label{fig:resolution:mom} Momentum resolution as a function of the full momentum}
\end{figure}

Figures~\ref{fig:resolution:dcax}, \ref{fig:resolution:dcay} and \ref{fig:resolution:dcaz} show the measured resolution for the distance of closest approach (DCA) to the primary vertex as a function of the transverse momentum of the particle. Figure~\ref{fig:resolution:mom} shows the momentum resolution. The figures demonstrate good agreement for the DCA resolution between our model and the detailed simulation. For momentum resolution, the agreement is good for $p>0.9$\,GeV, while at lower momentum values our model predictions result in a slightly overestimated resolution, compared to the detailed simulation.

\begin{figure*}
\centering
\begin{subfigure}{0.49\textwidth}
\centering
\includegraphics[width=\textwidth]{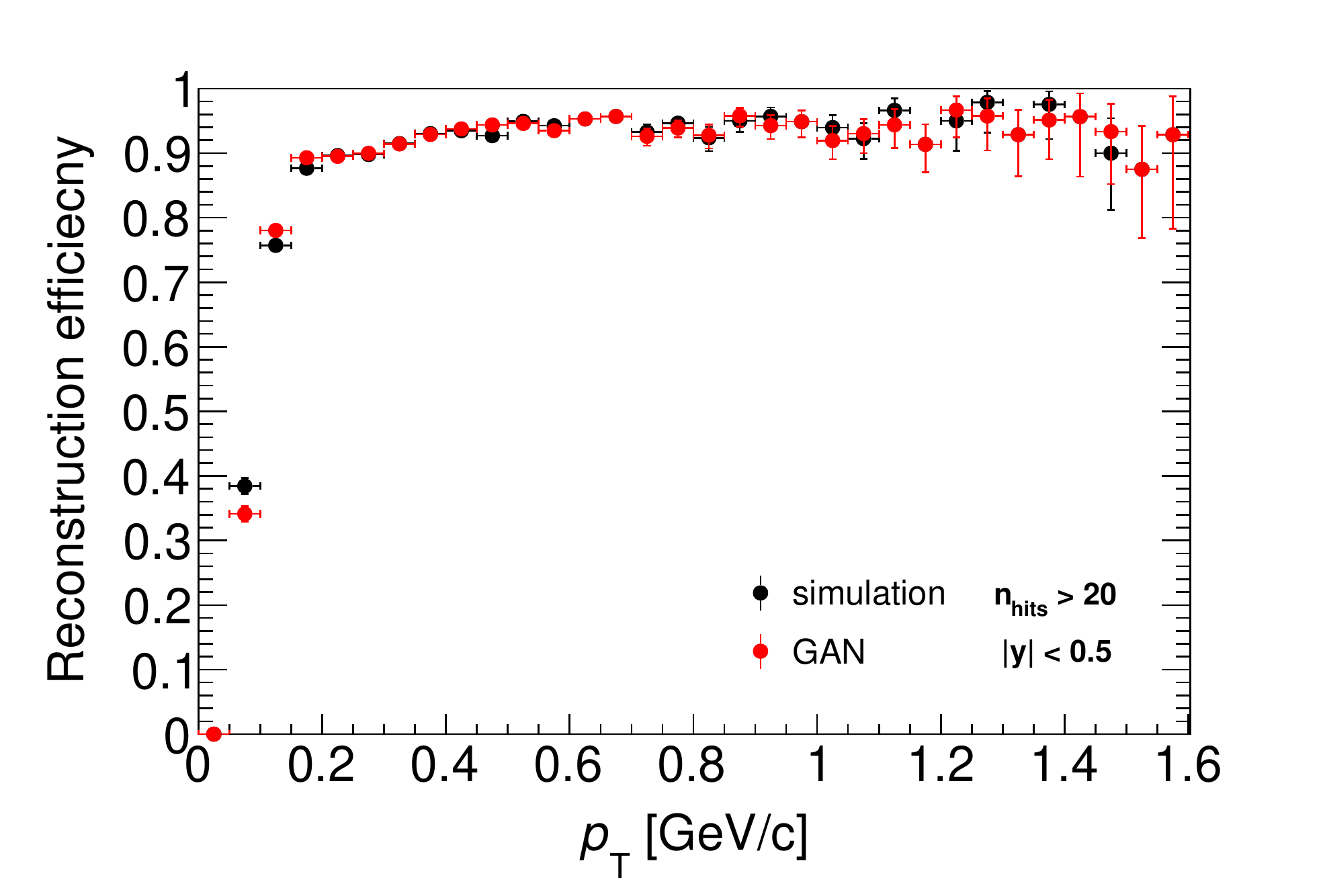}
\end{subfigure}
\begin{subfigure}{0.49\textwidth}
\centering
\includegraphics[width=\textwidth]{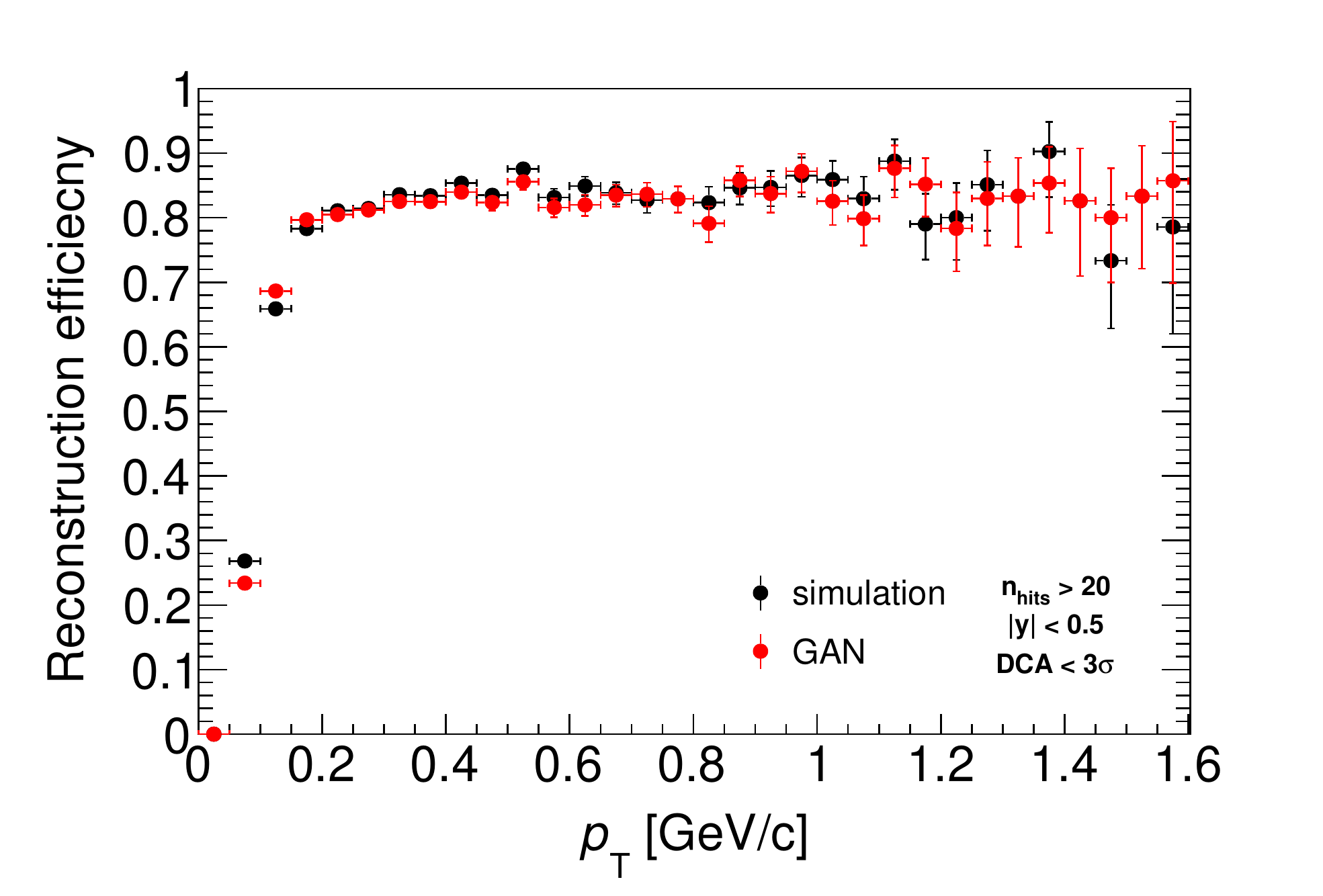}
\end{subfigure}
\\
\begin{subfigure}{0.49\textwidth}
\centering
\includegraphics[width=\textwidth]{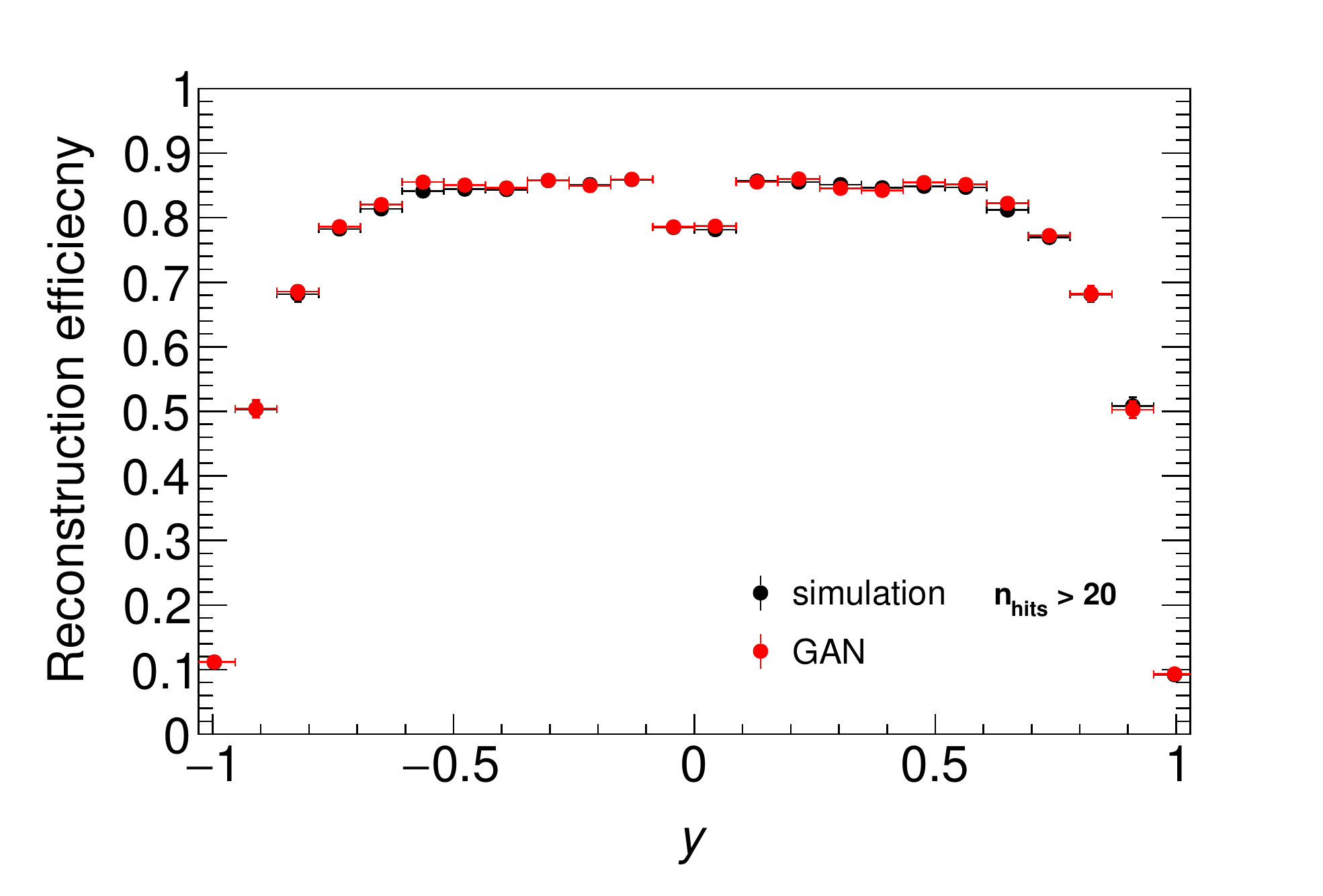}
\end{subfigure}
\begin{subfigure}{0.49\textwidth}
\centering
\includegraphics[width=\textwidth]{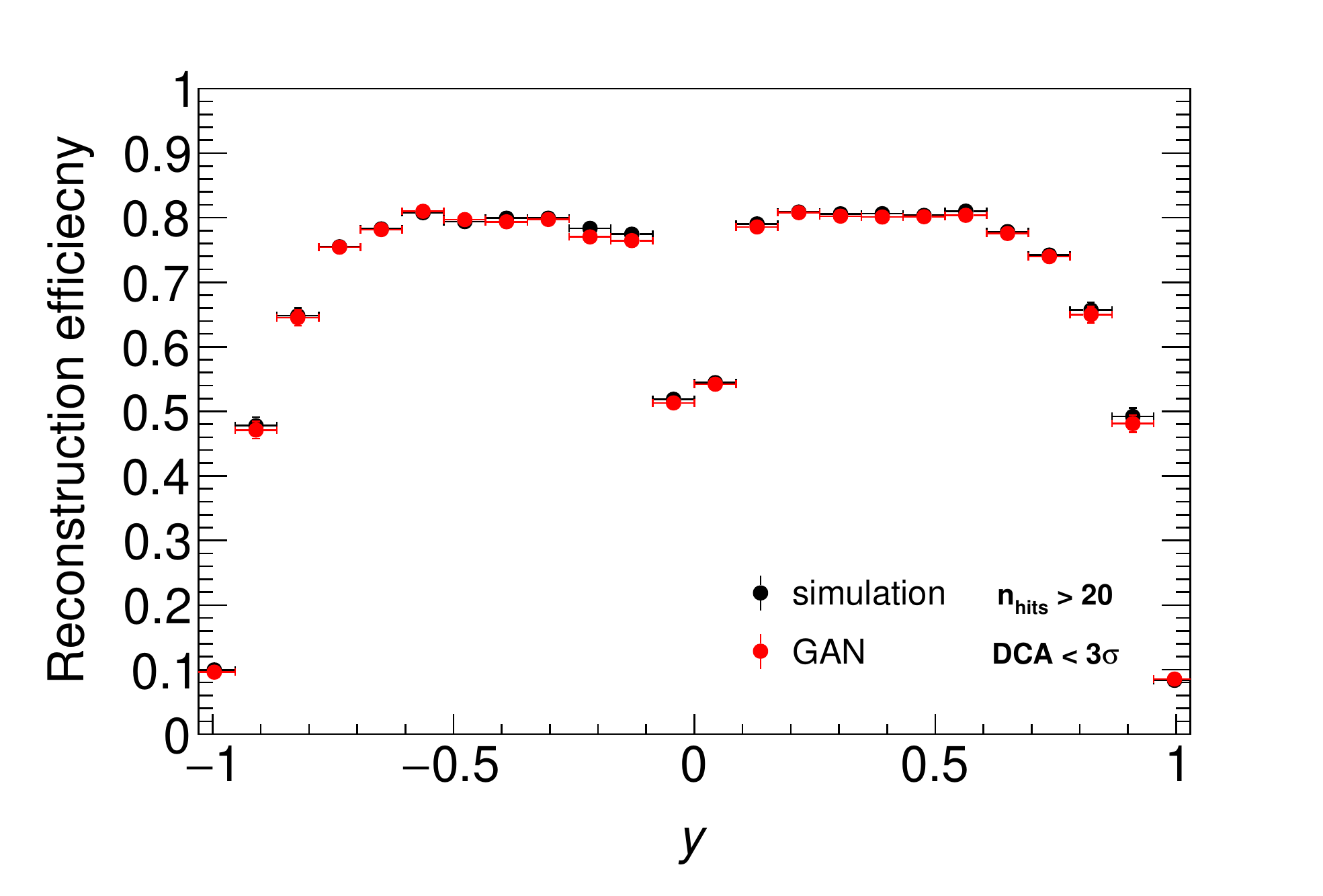}
\end{subfigure}
\caption{\label{fig:recoeff} Reconstruction efficiency as a function of the transverse momentum~(top row) and rapidity~(bottom row). Right column corresponds to the additional requirement on the reconstructed tracks quality~($\text{DCA}<3\sigma$).}
\end{figure*}

Figure~\ref{fig:recoeff} shows the track reconstruction efficiency as a function of the transverse momentum and rapidity. This quantity is shown for regular and tight track selection criteria, the latter corresponding to an additional requirement on the track DCA. These plots demonstrate reasonable agreement between our model and the detailed simulation in reconstruction efficiencies.

\begin{figure*}
\centering
\begin{subfigure}{0.33\textwidth}
\centering
\includegraphics[width=\textwidth]{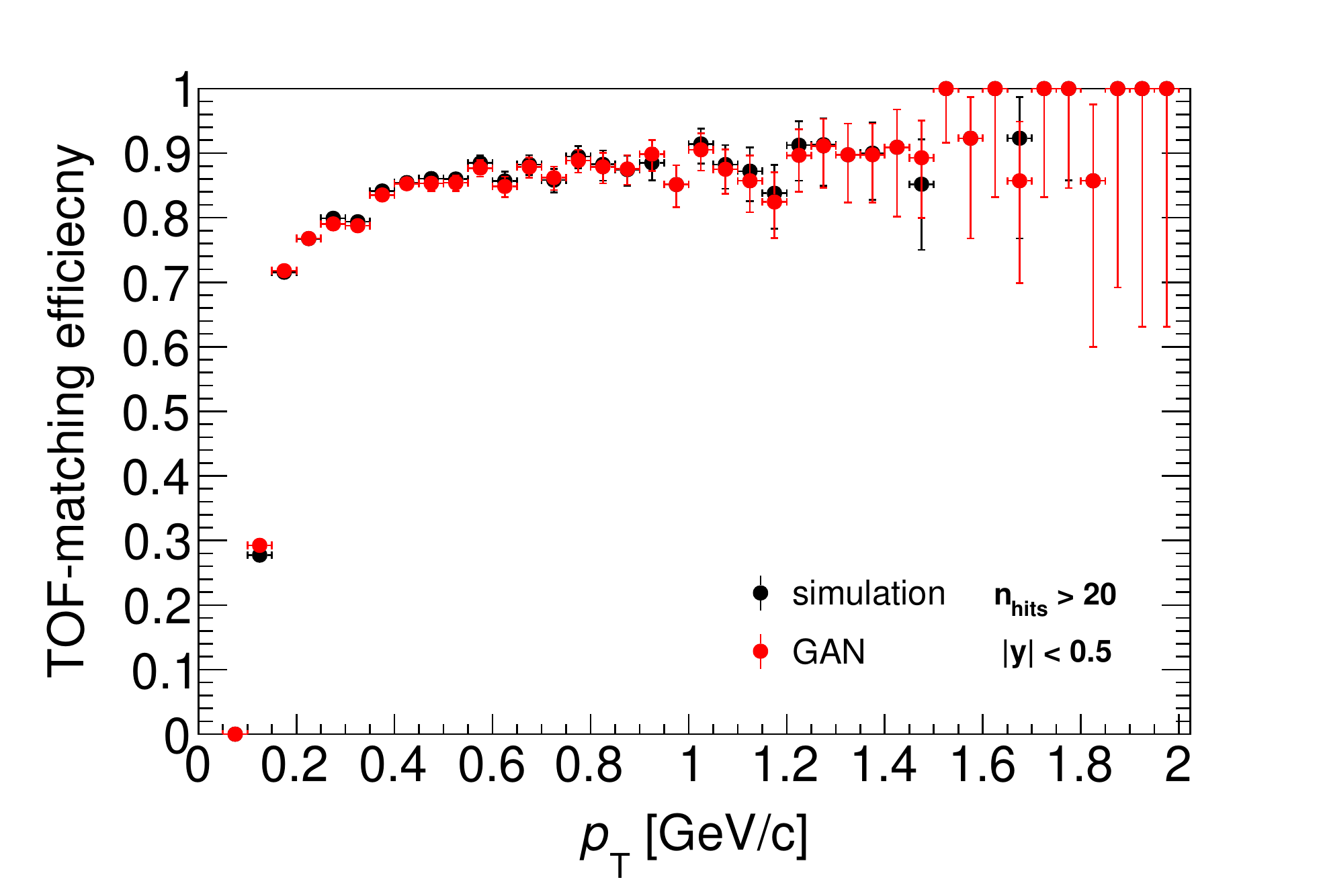}
\caption{\label{fig:tofmatchingpt} TOF matching efficiency as a function of the transverse momentum}
\end{subfigure}
\begin{subfigure}{0.33\textwidth}
\centering
\includegraphics[width=\textwidth]{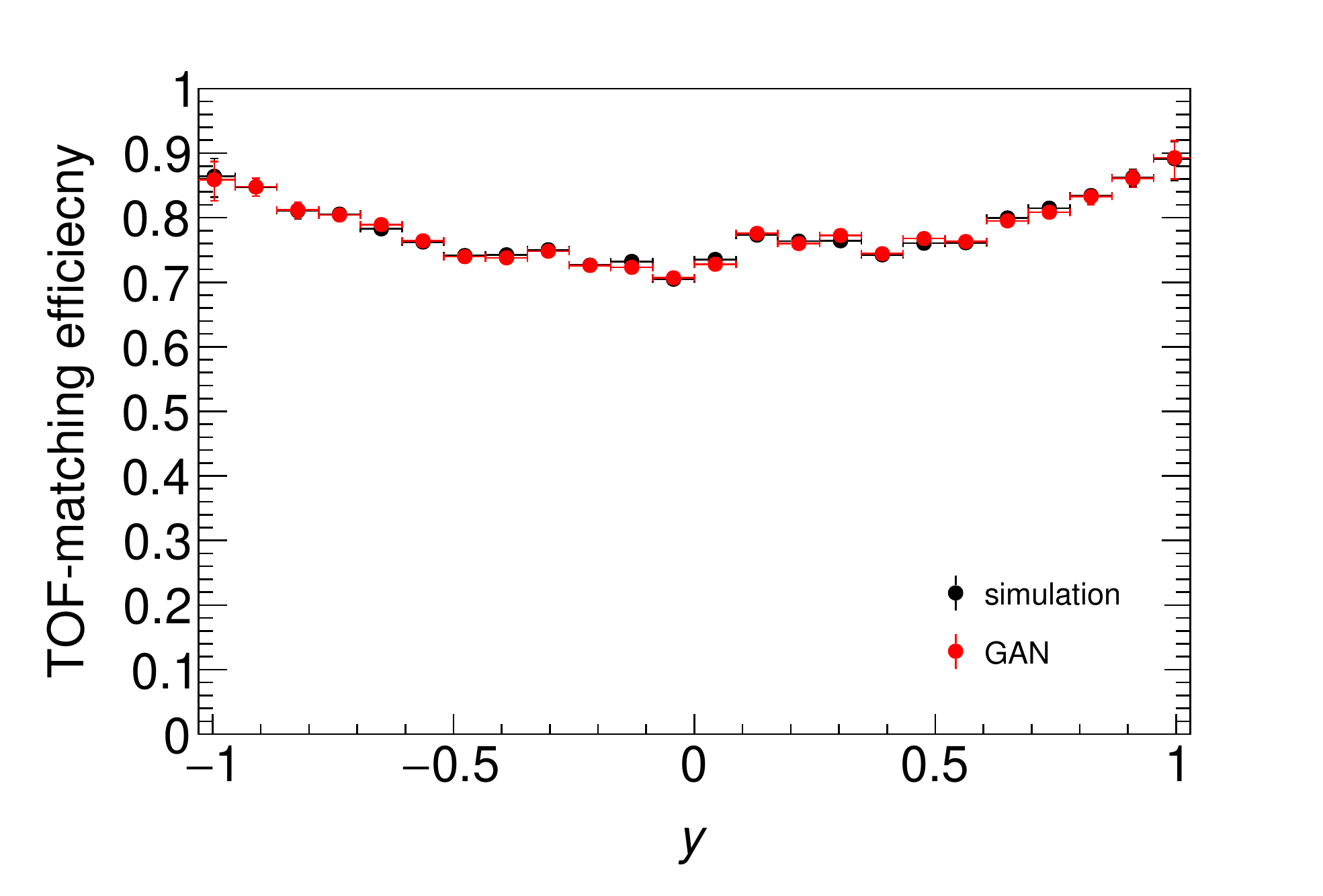}
\caption{\label{fig:tofmatchingrap} TOF matching efficiency as a function of the rapidity}
\end{subfigure}
\begin{subfigure}{0.33\textwidth}
\centering
\includegraphics[width=\textwidth]{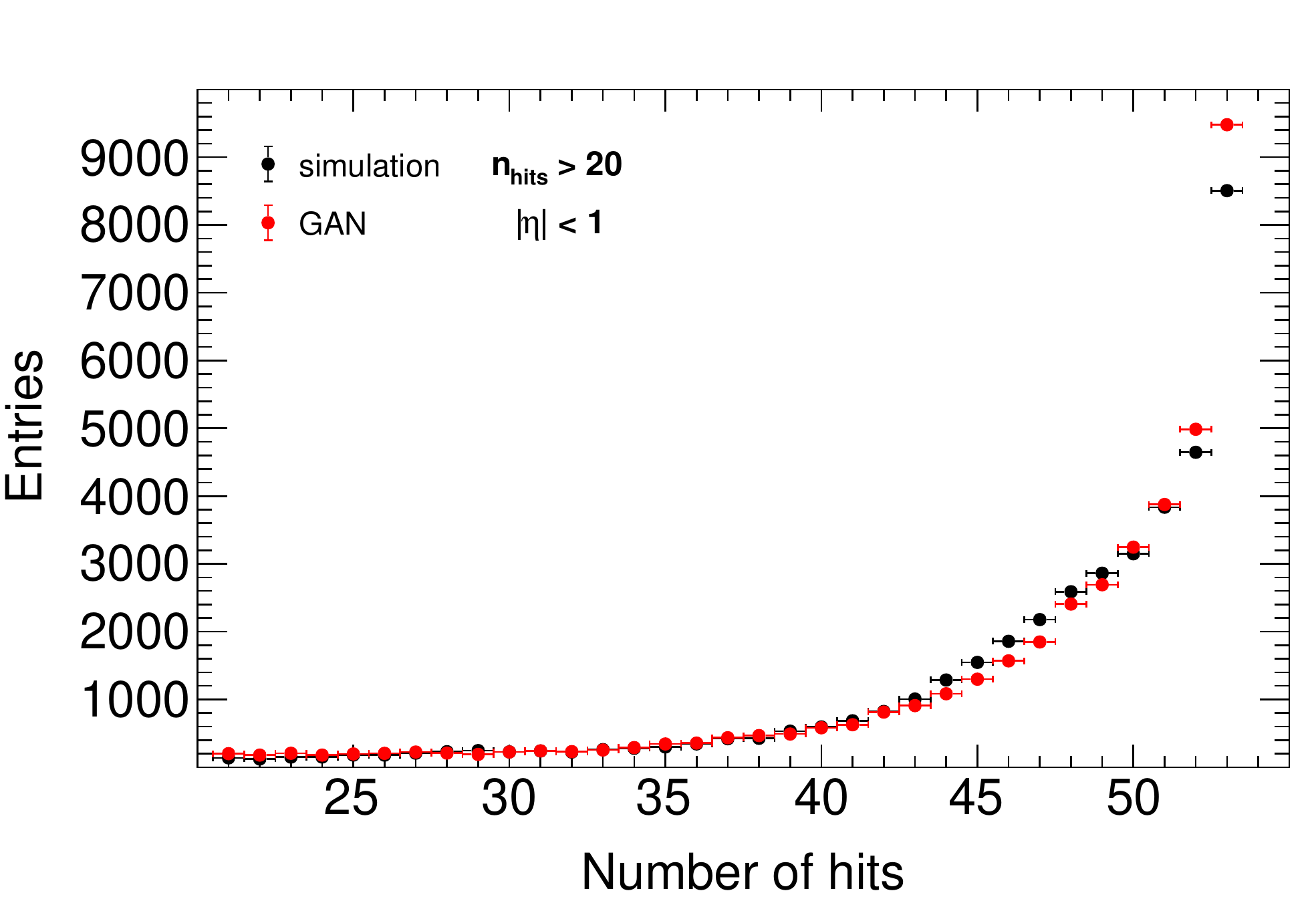}
\caption{\label{fig:nhits} Distribution of the number of hits per track}
\end{subfigure}
\caption{\label{fig:tofmatchingandnhits} TOF matching efficiencies over transverse momentum~(\subref{fig:tofmatchingpt}) and rapidity~(\subref{fig:tofmatchingrap}) and distribution of the number of hits per track~(\subref{fig:nhits})}
\end{figure*}

Finally, Fig.~\ref{fig:tofmatchingandnhits} shows the efficiency of matching the tracks to the signals from the Time-of-Flight (TOF) system of the MPD detector as a function of the transverse momentum and rapidity (Figs.~\ref{fig:tofmatchingpt} and~\ref{fig:tofmatchingrap}, respectively), and the distribution of the number of hits on track~(Fig.~\ref{fig:nhits}). The demonstrated agreement is excellent for the TOF matching efficiency, while the number of hits distributions are slightly inconsistent, with our model resulting in a slightly larger number of hits measured for a track. This effect is consistent with the overestimated momentum resolution and happens, as we believe, because the model is trained on the data from only the short TPC pads while utilized for the whole detector.

\begin{figure}
    \begin{minipage}{\columnwidth}
    \centering
    \includegraphics[width=\textwidth]{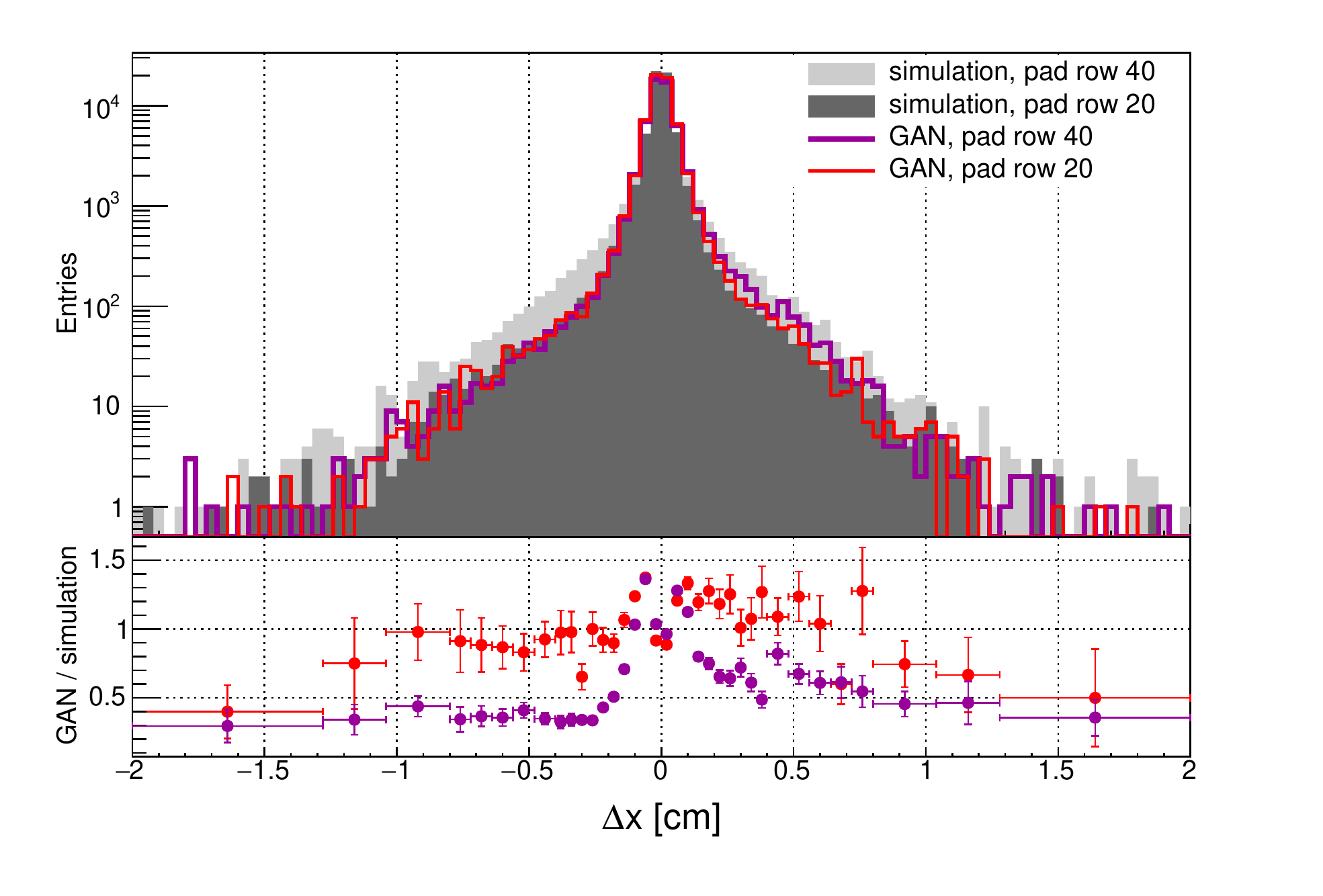}
    \end{minipage}
    \caption{\label{fig:dx} Distributions of differences $\Delta x=x_\text{reconstructed}-x_\text{true}$ between the reconstructed and true cluster coordinates along the pad row direction. For the short (long) pads from the pad row 20 (40), the detailed simulation results are shown in the dark (light) gray shaded histogram, while the histogram for the GAN prediction is shown with the red (magenta) line. The ratios between the GAN and detailed simulation yields in the same pad rows are shown in the bottom part of the plot.}
\end{figure}

To demonstrate that not taking into account the difference between the long and short pads may result in the observed discrepancies, we plot distributions of deviations $\Delta x=x_\text{reconstructed}-x_\text{true}$ of the reconstructed from the true cluster coordinates for rows of short (pad row 20) and long (pad row 40) pads in Fig.~\ref{fig:dx}, where $x$ is the coordinate along the pad row direction. This should reflect the coordinate resolution of the pads. As one would expect, the GAN predictions are similar for both short and long pads, and in a reasonable agreement with the detailed simulation results for the short pads, with slight inconsistencies in the shape in the center of the peak and far in the tails. The coordinate resolution, however, is worse for the long pads, as is predicted by the wider $\Delta x$ distribution from the detailed simulation, which is not captured by the GAN.

\section{Discussion}
\label{sec:discussion}
As was shown in section~\ref{sec:resultsandvalidation}, the proposed model demonstrates good performance over most of the metrics considered. The barycenters and widths of the pad response distributions are well reproduced. Track reconstruction and TOF-matching efficiencies, as well as vertexing resolution, agree between the GAN and the detailed TPC simulation model predictions. We observe good agreement in a wide kinematic range, although the model is originally trained on responses from pions with a fixed transverse momentum. This can be explained by the fact that momentum and particle type mostly affect the signal amplitude rather than the shape. While the former is important for the $dE/dx$ measurements, it is the latter that defines tracking characteristics. In fact, momentum does affect the shape of the signal through the curvature of the track, though this should be a rather little effect since the pad size is much smaller than the radius of the curvature.

Our model predictions result in an overestimated momentum resolution which goes in line with predicting more hits on track. These biases are likely to be caused by training our model on the responses from only the short pads (that have better coordinate resolution compared to the long ones) while making predictions for the whole detector. The integrated response amplitude distributions, though being reproduced well enough on average, are captured in the GAN with a slightly lower spread compared to the validation data. This may have an effect on the $dE/dx$ measurements in TPC and would require further tuning of the GAN model, or even modifications to the architecture, e.g. a separate factorized model to only predict the integrated amplitude. Alternatively, one can use Geant3 energy deposits in the detector material to scale the predicted integrated amplitude and therefore account for any mismodeling of the amplitude introduced by the GAN. Investigation of the $dE/dx$ performance, however, is beyond the scope of this work.

Along with possible enhancements in the amplitude modeling and incorporating the pad type into our model, further developments could introduce various particle types and momentum of the particle at a given track segment as additional input parameters. It is also important to evaluate the bias introduced by factorizing the responses at the adjacent pad rows. 

To evaluate the performance speed-up, we run the detailed and fast TPC models on a single core of an Intel Core i7-3770K (3.50GHz) CPU, with no GPU acceleration. These tests show the GAN model integrated into the MPD software running 12 times faster compared to the detailed simulation on the central Au+Au events.

\section{Conclusion}
\label{sec:conclusion}
In this work, we have shown a\RemoveText{n original}\AddText{~new} approach to fast simulation of TPC type detectors, based on a Generative Adversarial Network. Our model is built for the TPC detector of the MPD experiment at the NICA accelerator complex. In our approach, we split the charged particle tracks into small segments contributing to different rows of the sensitive TPC pads, and then generate the pad responses at a given row conditioned by the track segment parameters. A custom activation function is used at the generator output layer to account for the discrete mode of the response distribution, caused by the noise threshold in the detailed simulation.

We have evaluated the predictions of our model by both comparing elementary response distribution characteristics, i.e. 1st and 2nd order moments and integrated amplitudes, as well as the reconstructed event properties like vertex and momentum resolution and track reconstruction efficiencies. Most of the evaluation metrics show good agreement of our model with the detailed simulation. The mismodeling of the variance of the integrated amplitude can be mitigated by scaling it with the Geant3 energy deposits in the detector material. The few inconsistencies that are seen in the reconstructed characteristics may be attributed to training the model on the responses from the short pads only while using it to predict on both short and long pads. We expect our model to be ready for use in physics tasks once trained with transverse momentum and pad type information taken into account.

Integrated into the MPD software, the proposed model runs 12 times faster on central Au+Au events, compared to the detailed simulation. This speedup translates to a factor of 2 improvement in the time spent on the overall simulation \& reconstruction pipeline and therefore is sufficient at the current stage of the MPD software development.

\begin{acknowledgements}
Contribution of A. Maevskiy, including development, training and evaluating the deep learning model, was supported by the Russian Science Foundation under grant agreement n$^{\circ}$ 19-71-30020.
Contribution of F. Ratnikov, including the coordination of the work and the quality metric studies, was supported within the framework of the Academic Fund Program at the National Research University Higher School of Economics (HSE) in 2020-2021 (grant n$^{\circ}$294715) and within the framework of the Russian Academic Excellence Project "5-100".
Contribution of A. Zinchenko, including Monte Carlo simulation and performance testing of the MPD TPC, was supported by the Russian Foundation for Basic Research under grant agreement 18-02-40060.
This research was supported in part through computational resources of HPC facilities at NRU HSE.
\end{acknowledgements}

\appendix
\section{Network architecture}
\label{sec:appendix:nnarchitecture}
\begin{figure*}[p]
\centering
\includegraphics[width=0.49\textwidth]{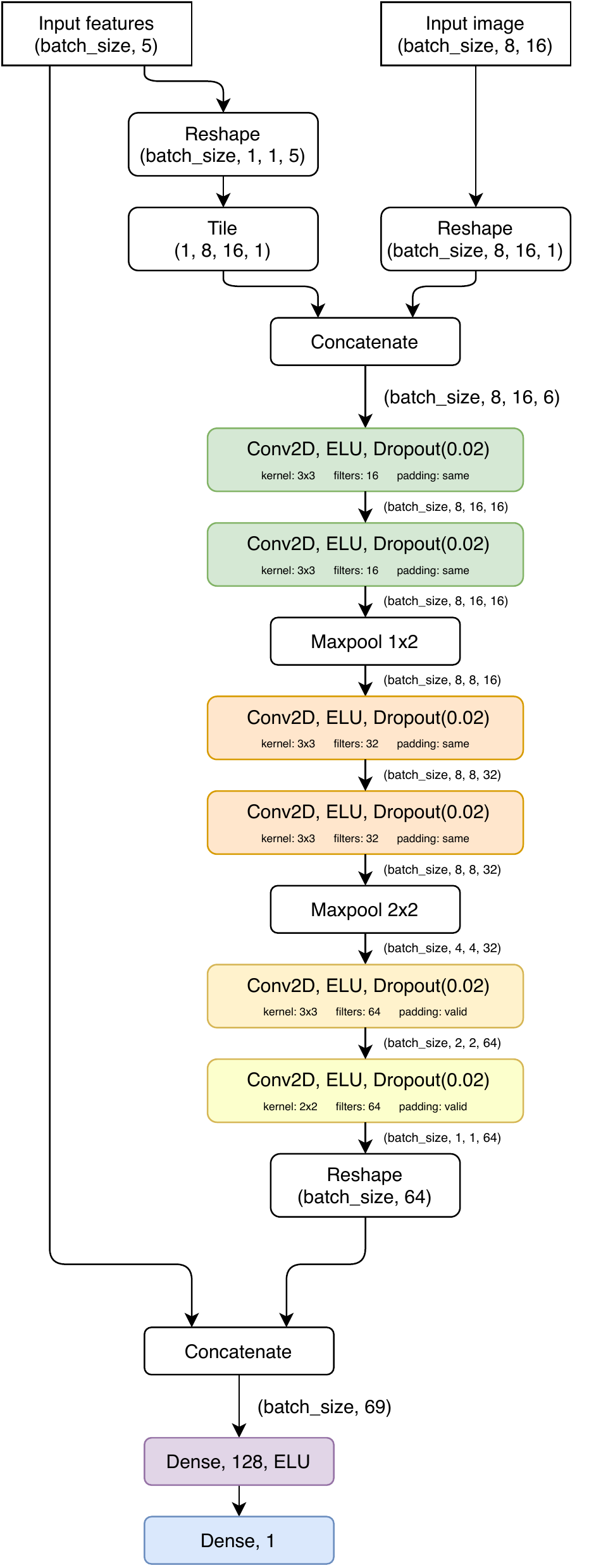}
\hfill
\raisebox{.5\height}{\includegraphics[width=0.49\textwidth]{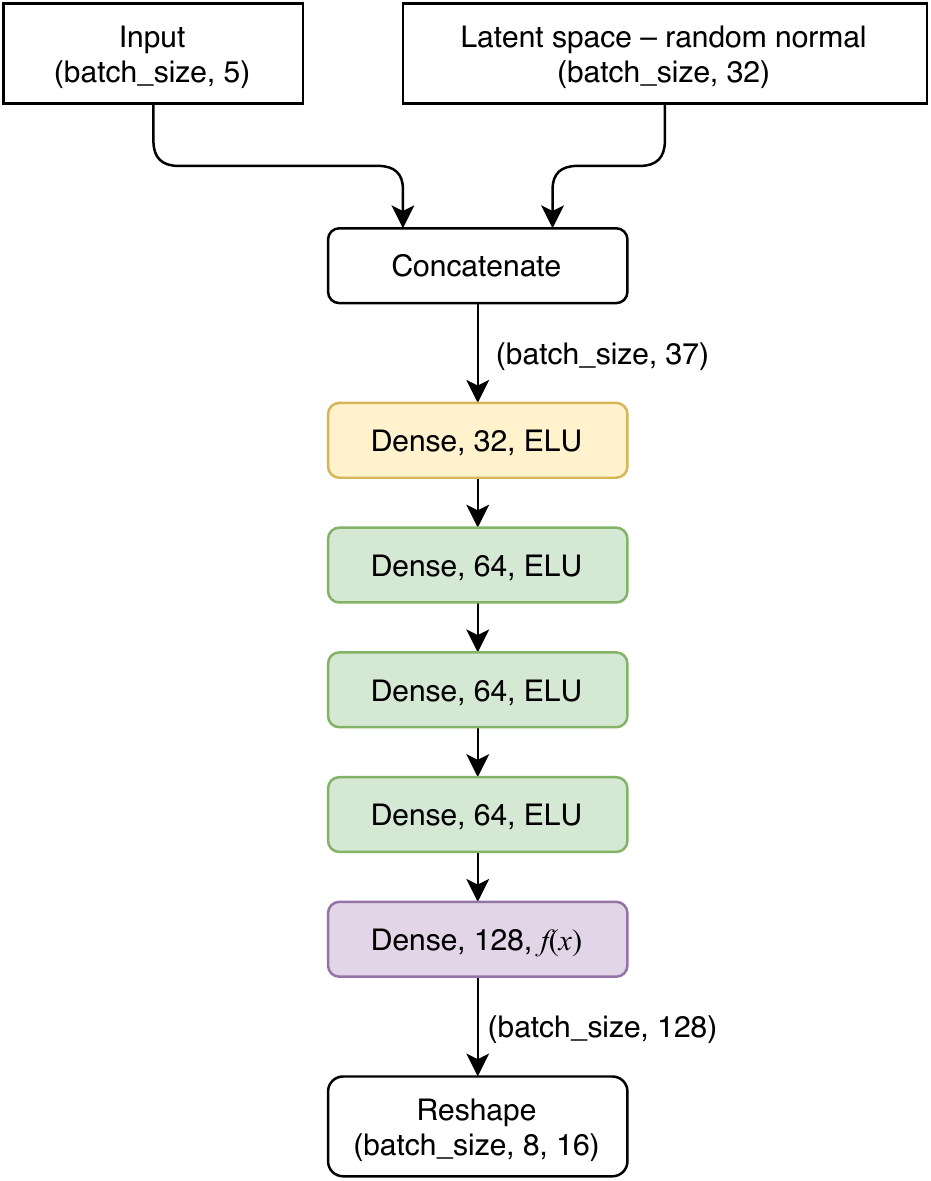}}
\caption{\label{fig:architecture} Neural network architecture for the discriminator (left) and the generator~(right)}
\end{figure*}

The neural network architectures for the discriminator and generator are shown in Fig.~\ref{fig:architecture}.

\providecommand{\doi}[1]{\href{https://dx.doi.org/#1}{DOI \discretionary{}{}{}\ReplaceStr{#1}}}
\small
\bibliographystyle{spphys}
\bibliography{references}

\end{document}